\documentstyle[aps]{revtex}
\newcommand{\myfig}[3]{
        \begin{figure}[ptb]
        \caption[#2]{#3}
        \label{#1}
        \end{figure} }

\begin{document}

\title{Triangular Antiferromagnets} 

\author{M. F. Collins}
\address{Department of Physics and Astronomy, McMaster University, 
Hamilton, Ontario L8S 4M1, Canada \\
E-mail: mcollins@mcmail.cis.mcmaster.ca \\
Tel: 905-529-7070, ext. 24172 \\
FAX: 905-546-1252 } 

\author{O. A. Petrenko}
\address{Department of Physics, University of Warwick,
Coventry, CV4 7AL, United Kingdom \\
E-mail: phsby@csv.warwick.ac.uk \\
Tel: +44 (1203) 523-414 \\
FAX: +44 (1203) 523-603}

\maketitle

\vspace*{13mm}

\begin{abstract}
In this article we review the effects of magnetic frustation in the stacked 
triangular lattice. Frustration increases the degeneracy of the ground 
state, giving rise to different physics. In particular it leads to unique phase 
diagrams with multicritical points and novel critical phenomena. We 
describe the confrontation of theory and experiment for a number of systems 
with differing magnetic Hamiltonians; Heisenberg, Heisenberg with easy-axis 
anisotropy, Heisenberg with easy-plane anisotropy, Ising and singlet 
ground state. Interestingly each leads to different magnetic properties and 
phase diagrams. We also describe the effects of ferromagnetic, rather than 
antiferromagnetic, stacking and of small distortions of the triangular 
lattice.

\end{abstract}

\newpage

\section{Introduction} 
\label{Intro}

Although  very  few  magnetic  systems   can  be  solved  exactly  in   three
dimensions, there  is a  reasonable understanding  of the  physics of  simple
systems  without  competing  interactions.  Numerical  estimates for physical
quantities exist to at least the accuracy that experiments currently  attain.
Many systems do exhibit competing interactions however, that is  interactions
that  do  not  all  favour  the  same  ordered  state.  Sufficiently   strong
competition can lead to new physics which is manifested by the appearance  of
non-collinear ordering, novel critical exponents, rich phase diagrams, or  an
absence of long range order at low temperatures.

There  are  several  ways  in  which  frustration  can arise. In this article
attention  is  limited  to  geometric  frustration  arising  from  triangular
arrangements    of    magnetic    moments    with    each    pair     coupled
antiferromagnetically.  Figure~1a shows such a situation  with
three atoms forming an  equilateral triangle. Atoms 1  and 2 form a  state of
lowest energy when  their moments are  aligned antiparallel, but  then atom 3
cannot align itself simultaneously antiparallel  to the moment on both  atoms
1 and 2, so  it is frustrated. This  frustration is most severe  for an Ising
system where  spins can  align only  in one  direction. For  classical vector
spins confined  to the  plane of  the paper  (XY interactions)  there are two
degenerate solutions for  the lowest energy  of the system  for a given  spin
vector  at  atom~1.  These  are  shown  in  figures~1b and 1c.  
Both have the three spin vectors at \(120^\circ\) to  each
other, and  the degeneracy  corresponds to  two different  chiral states. The
overall degeneracy  of the  ground state  involves the  product of this state
and  orientation  of  spin~1  in  the  plane;  this corresponds to the group
\(Z_{2}{\bf \times}S_{1}\). For three dimensional vector spins and  isotropic
Heisenberg interactions  the degeneracy  is greater  still since the spin 
on atom 1 can be in any direction and then the spins on atoms 2 and 3 
can be in any plane which contains the direction of spin 1.This degeneracy 
corresponds to the group SO(3). If the Hamiltonian has easy axis 
anisotropy, the plane of the three spins will contain the Z axis and the 
angle between the three spin vectors will no longer be 120$^\circ$. A 
general feature of these frustrated systems  is that the  ground
state  has  extra  degeneracy  over  and  above  that  found in the analogous
non-frustrated systems.  This is what  gives rises to the possibility of  new
physics.

There are several ways  that triangles of antiferromagnetic  interactions can
be built into a  crystal lattice. In this  article we limit consideration  to
the effects  of frustration   in the  stacked triangular  lattice. This  is a
lattice   containing   two-dimensional    triangular   sheets   coupled    by
antiferromagnetic   interactions   $J'$,   with   the   sheets   stacked
perpendicular  to  the  plane  with  coupling $J$ between  neighbouring atoms
in different planes. Almost  all the work on this lattice involves one of two
crystal structures,  one with  composition ABX$_3$  where A  is an alkali
metal, B is a  transition metal, and X  is a halogen atom,  and the other
with  composition  BX$_2$. The crystal structure of these materials is 
shown in figure 2. .
 In  ABX$_3$ compounds  there  are  chains of
magnetic  B  atoms   along  the  Z   direction  coupled  by   superexchange
 interactions  through  three  equivalent  anions  X.  There  is  no direct
 superexchange coupling between atoms on neighbouring chains. The  magnitude
 of the  interplanar interactions  $J$ is  typically two  to three  orders of
magnitude  greater  than  intraplanar  interactions  $J'$,  so that at high
temperatures  the  magnetic  properties become  quasi  one  dimensional.  In
BX$_2$ compounds  the strong  superexchange coupling  is intraplanar
  and  the  ratio  of  the  magnitude  of  the  interactions is
reversed.  At  high  temperatures  the  magnetic  properties  are  quasi  two
dimensional. 
In a third known type of triangular antiferromagnets, ABO$_2$
compounds, the stacking of two-dimensional triangular sheets along the Z 
direction is different from ABX$_3$ and BX$_2$ compounds, the stacking 
sequence is rhombohedral of ABCABCABC... type. The crystal structure dictates 
strong two dimensional character of magnetic system in ABO$_2$ compounds: 
the exchange path of the interplane interaction $J'$, B--O--A--O--B, is 
much longer than that of the intraplane interaction $J$, B--B and B--O--B. 
At low temperatures almost all mentioned above materials form
three-dimensional ordered magnetic structures  that indicate the presence  of
magnetic frustration  in the  triangular lattice.  In many,  though not  all,
cases  the  low  temperature  structure  is  built from one of the structures
shown in figure~1. 

Kawamura   \cite{t_Kawamura_87,t_Kawamura_89}   predicted   that   the  extra
degeneracy in the  ground state of  triangular antiferromagnets leads  to new
physics, which  can be  described in  terms of  universality classes based on
the  symmetry  of  the  order  parameter.   Monte Carlo 
work~\cite{Kawamura_92,Bhattacharya_94,Loison_94,Mailhot_94} supports Kawamura's 
prediction for Heisenberg symmetry as does 4-$\epsilon$ renormalization group 
calculations \cite{Kawamura_88}.  However the non-linear $\sigma$ model in 
2+$\epsilon$ dimensions\cite{t_Azaria_90} indicates non-universal behavior, 
likely with mean-field tricritical exponents, and 3D renormalization group 
calculations with a resummation technique show a first-order phase 
transition~\cite{Antonenko_94}.  The situation is no less clear  for XY symmetry.

Experimentally  stacked  triangular  antiferromagnets  are found to have 
critical phase transitions with critical exponents that do not  correspond to
any of the standard universality classes for both Heisenberg
\cite{vcl2_Kadowaki_87} and XY
\cite{cmb_Mason_89} symmetry.  In fact  the critical  points are  found to be
tetracritical in character \cite{cmb_Gaulin_89}  so that they clearly  cannot
belong in  the same  universality class  as unfrustrated  systems. This 
shows that the physics of magnetic systems can be changed by frustration.
Both experiment and theory indicate that the ordered states found at low 
temperatures in frustrated systems have reduced ordered moments.  Any detailed 
theory must therefore include quantum effects in determining the ground state, 
and classical theory may be substantially in error.

In fact much of the physics of the stacked triangular lattice is similar to 
that predicted for the two-dimensional triangular 
lattice~\cite{t_Lee_84,t_Chubukov_91,t_Momoi_94} and the presence of the third 
dimension is not an essential ingredient of most of the ideas that are used 
in the field. However, for all well-characterized materials which have triangular 
magnetic lattices, the long-range order at low temperatures is three-dimensional 
in nature. For nearest-Heis.texneighbour ferromagnetic or antiferromagnetic 
interactions, there is no extra frustration involved by going to three dimensions, 
and the main influence of the presence of the third dimension is that the 
two-dimensional ordering process is stabilized.
 
In  this  article  we  describe  in  some  detail  the magnetic properties of
materials  with  stacked  triangular  lattices,  paying  special attention to
cases where  there are  novel physical  phenomena absent  in the unfrustrated
case. Emphasis is placed on  the experimental results, though where  possible
the discussion  is put  in the  appropriate theoretical  context. An additional 
information about triangular antiferromagnets may be found in the related 
reviews:~\cite{Gaulin_94,Plumer,Ramirez,Hatfield}.
We classify the stacked triangular lattice materials by the nature of the
magnetic Hamiltonian. Chapter~\ref{Heis} describes materials with Heisenberg
interactions, Chapter~\ref{Ising} describes systems with Ising Hamiltonians,  
Chapter~\ref{SGS} describes singlet ground state magnets and 
Chapter~\ref{Ferro} considers cases with ferromagnetic stacking of the planes.
Chapter~\ref{Impurity} describes magnetic properties of the diluted triangular 
antiferromagnets. Chapter~\ref{Concl}  presents conclusions from these studies.

We conclude this section by listing  the materials that have been studied  in
the  context  of  stacked  triangular  antiferromagnetism. 

a) Table  \ref{listofmagnstr}  shows  the  magnetic  structure  of  ABX$_3$
triangular compounds. The cases where B is a chromium atom or A is a thallium 
atom are omitted since these compounds do not form with a triangular lattice.

b) Four BX$_2$ have been studied: VCl$_2$ and VBr$_2$ are close to Heisenberg 
systems~\cite{vcl2_Kadowaki_87}, though there is weak easy-axis anisotropy, 
the magnetic structure of VI$_2$ is not clear~\cite{vi2_Kuindersma_79,vx2_Hirakawa_83}. 
MnBr$_2$  has a  complex 
magnetic structure which is not triangular in nature~\cite{mb_Sato_94}.

c) Among  ABO$_2$ compounds magnetic properties of only three triangular
antiferromagnets have been investigated in details. LiCrO$_2$ and CuCrO$_2$ 
demostrate 120$^\circ$ magnetic structure with weak easy-axis
anisotropy~\cite{lco_Soubeyroux_81,lco_Kadowaki_95,cco_Kadowaki_90},  
AgCrO$_2$ has slightly modulated 120$^\circ$ structure~\cite{aco_Oohara_94}. 

\section{Heisenberg triangular antiferromagnets}
\label{Heis}

The stacked triangular magnetic lattice is shown in figure~3. 
Its magnetic properties for Heisenberg-type antiferromagnetism 
with single-ion anisotropy can be descibed on the basis of the 
following Hamiltonian:
\large 
\begin{equation}
\hat{\cal H} = J\!\sum_{i,j}^{\text{chains}}\!{\bf S}_i{\bf S}_j +
J'\!\sum_{k,l}^{\text{planes}}\!{\bf S}_k{\bf S}_l + 
D\!\sum_i(S_i^z)^2 - g \mu_B {\bf H}\sum_i{\bf S}_i,
\label{H_H}
\end{equation} 
\normalsize
where ${\bf S}$ is a spin of the magnetic ion, $J$ is the exchange 
integral along the $c$ axis of the crystal, $J'$ is the exchange integral 
in the perpendicular direction, $D$ is the anisotropy constant, whose sign 
determines the orientation of the spin plane relative to the crystal axes. 
The first sum describes the exchange energy along the chain, the second sum 
describes the exchange energy in the basal plane, and the third and fourth 
sums represent the single-ion anisotropy energy and the Zeeman energy of the 
spins in a external magnetic field  ${\bf H}$ respectively. 
The case $D=0$ corresponds to a pure Heisenberg system. All real triangular 
magnets have non zero $D$, but if $|D|$ is small compared with both $|J|$ 
and $|J'|$ the magnetic properties will be close to those of Heisenberg 
systems except at very low temperatures $T<DS^2$ or very close to the 
critical point. If $D>0$ the ground state in zero field has spins confined 
to the $XY$ plane and in the critical region the fluctuations will only 
diverge for spin components within the $XY$ plane. If $D<0$ the anisotropy 
energy will be minimised for spins aligned perpendicular to the $XY$ plane. 
This term will compete with the
antiferromagnetic $J'$ exchange term and lead to additional frustration. 

Table \ref{Heis_parameters} lists the parameters $J$, $J'$ (both assumed 
nearest neighbour interactions only) and $D$ that have been determined 
experimentally in units of frequency, or energy divided by $h$.  The data 
confirm our previous observation that ABX$_3$ compounds are 
quasi-one-dimensional with $J{\gg}J'$ and that VX$_2$ compounds are 
quasi-two-dimensional with $J{\ll}J'$. In the ABO$_2$ compounds only the 
in-plane exchange, $J'$ has been measured reliably.

\subsection{Quasi-two-dimensional triangular antiferromagnets of type 
BX$_2$ and ABO$_2$}

VCl$_2$ and VBr$_2$ crystallize in the CdI$_2$ structure with a space 
group P\=3m1. VI$_2$ exists in two modifications, black and red. The 
black modification is composed of a statistical alternating layer 
structure of the CdI$_2$ and the CdCl$_2$ type, while the red modification 
crystallizes in the CdI$_2$ structure \cite{Juza_69}. This feature may 
explain the fact that the majority of work on VI$_2$ was done on powder, 
rather than on single crystal -- it is difficult to prepare good quality 
single crystal of VI$_2$. 

VCl$_2$ and VBr$_2$ are important because they are the stacked triangular
materials with Hamiltonians closest to the Heisenberg form. Both materials 
show critical phase transitions from ordered to paramagnetic states in zero 
field \cite{vcl2_Kadowaki_87,vbr2_Takeda_86,vbr2_Wosnitza_94}.

Despite numerous efforts to determine magnetic structures of BX$_2$ 
antiferromagnets, including NMR spectra and relaxation measurements 
\cite{vcl2_Tabak_93}, ESR measurements \cite{vx2_Yamada_84} and direct 
neutron scattering 
measurements~\cite{vi2_Kuindersma_79,vx2_Hirakawa_83,vbr2_Nishi_84,vbr2_Kadowaki_85,vcl2_Kadowaki_87}  
there are still some open question about spin-structures. Neutron polarization 
analysis  has shown that below $T_N$ the spin structure of VCl$_2$ is the 
120$^\circ$ structure in the $ac$-plane \cite{vcl2_Kadowaki_87}, while in 
the case of VBr$_2$ neutron scattering results are consistent with both the 
120$^\circ$ structure in the $ac$-plane and a partially disordered structure 
whose spins cant from the $c$-axis by 45$^\circ$ 
\cite{vbr2_Nishi_84,vbr2_Kadowaki_85}. 
VI$_2$  has only been looked by neutron scattering at in powder form. 
Two neutron powder measurements \cite{vi2_Kuindersma_79,vx2_Hirakawa_83} 
give different patterns. Solution of the structure will probably need 
single-crystal data, the simple triangular structure is not observed. 
There are two phase transitions, a critical transition at 16.3~K and a 
first-order transition at 14.4~K \cite{vx2_Hirakawa_83}.
  
VCl$_2$ has weak easy-axis anisotropy as is shown by the data in 
table~\ref{Heis_parameters} and also by the splitting of the critical point 
from the single tetracritical point of the frustrated Heisenberg Hamiltonian 
into two ordinary critical transitions at  $T_{N1}=35.88(1)$~K and at 
$T_{N2}=35.80(1)$~K in zero field \cite{vcl2_Kadowaki_87}. This splitting  
of about one part in 450 will give rise to crossover behaviour in the critical 
region \cite{rev_Collins_89}. For reduced temperatures $t=(T-T_{N})/T_N$ 
of magnitude more than about 1/450 above $T_{N1}$ or below $T_{N2}$ the 
critical behaviour will be that of the frustrated Heisenberg system. Nearer to 
$T_{N1}$ or $T_{N2}$ the critical behaviour will reflect the fact that the 
correlation length only diverges along the $Z$ direction, not in the XY plane, 
and will be different. Thus it is possible in principle to measure three sets 
of critical exponents in VCl$_2$, though in practice only the Heisenberg 
exponents for $|t|>1/450$ have been measured. In VBr$_2$ the splitting of 
$T_N=28.66(2)$~K is even smaller and has not been observed 
\cite{vbr2_Wosnitza_94}. The measured critical exponents should correspond 
to those of the frustrated Heisenberg system.

Table \ref{critexp} lists the observed critical exponents in VCl$_2$ and 
VBr$_2$ and compares them with theoretical predictions. It is clear that the 
frustrated Heisenberg system has critical exponents that are far from those 
of the unfrustrated system, confirming that frustration changes the physics. 
Neither the SO(3) nor tricritical model are in complete accord with experiment; 
the SO(3) model is in disagreement with experiment for $\beta$ and $\gamma$ 
and the tricritical exponents are in disagreement for $\nu$ and $\alpha$. 
In every case both these models are significantly better than the 
unfrustrated Heisenberg model.

MnBr$_2$ also does not order in the triangular structure. Its structure, 
based on single-crystal neutron data, is complex based on arrangements with 
two up spins followed by two down spins \cite{mb_Sato_94}. There are two 
magnetic phase transitions: second order at $T_{N1}=$2.32~K and first order 
at $T_{N2}=$2.17~K.

LiCrO$_2$ has been studied by single crystal susceptibility and neutron 
diffraction measurements~\cite{lco_Soubeyroux_81,lco_Kadowaki_95}, optical 
measurements~\cite{vx2_LiCrO2_Kojima_93,lco_Suzuki_93}, 
ESR~\cite{abo_Angelov_84,hco_lco_Ajiro_88} and neutron diffraction
measurements~\cite{lco_Soubeyroux_79} on a powder sample. Three dimensional 
magnetic ordering, characterized by a double-$\bf Q$ 120$^\circ$ structure 
with non-equivalent wave vectors $\bf Q$ of $(\frac{1}{3} \frac{1}{3} 0)$ and 
$(-\frac{2}{3} \frac{1}{3} \frac{1}{2})$~\cite{lco_Kadowaki_95}, is observed 
below the single-phase-transition temperature $T_N=64$~K. Polarization analysis
of neutron scattering data shows that spins triangulars are confined in a plane 
including the $c$ axis, that is the magnetic anisotropy is of the easy-axis 
type. On the other hand, absence of splitting of $T_N$ and of anisotropy in the 
susceptibility above $T_N$ demonstrate that magnetic anisotropy $D$ is much 
smaller than exchange interactions $J$ and $J'$. No direct measurements of
$D$ and $J'$ have been reported, the only an attempt to estimate the
$J/J'$ ratio from the phase transition temperature has been made by 
Angelov and Doumerc~\cite{aco_Angelov_91}, which gives only a very rough 
estimate. The ratio $J/J'$ could also be estimated from susceptibility 
measurements since it is proportional to the 
$(\chi_{\parallel C}-\chi_{\perp C})/\chi$=5\% ratio (note the difference 
between LiCrO$_2$ and VX$_2$ compounds:
in LiCrO$_2$ $\chi_{\parallel C}>\chi_{\perp C}$). The optical and ESR 
measurements were mostly devoted to the
problem of finding of the characteristic point-defects known as $Z_2$ vortices,
predicted theoretically by Kawamura and Miyashita~\cite{t_Kawamura_84_2}.
It is believed the line width of the exciton magnon 
absorption~\cite{vx2_LiCrO2_Kojima_93} and EPR linewidth~\cite{hco_lco_Ajiro_88}
 exhibit $Z_2$-vortex induced broadening.

The rest of the ABO$_2$-type triangular antiferromagnets have been studied
only in a powder form. AgCrO$_2$ has been reported to order magnetically
below 24~K. It forms a slightly modulated 120$^\circ$ 
structure~\cite{aco_Oohara_94} with magnetic peaks at (0.327 0.327 0). 
The width of peaks indicates that the development of the true long-range
magnetic order is suppressed. Results of neutron powder diffraction studies
on CuCrO$_2$ below $T_N=(25\pm0.5)$~K are consistent with the 120$^\circ$ 
structure in the $a$-$c$ plane with moment 
$(3.1\pm0.2)\mu_B$~\cite{cco_Kadowaki_90}. No long range magnetic ordering 
has been found in NaTiO$_2$ and LiNiO$_2$ down to 
1.4~K~\cite{nti_lni_Hirakawa_85} and in NaCrO$_2$ and KCrO$_2$ down to 2~K
\cite{lco_Soubeyroux_79}.

An interesting issue is a magnetic moment reduction at low temperature. 
There are three general reasons which effect the value of magnetic moments 
in ABX$_3$ and VX$_2$ compounds: 1) covalency reduction, 2) effect of 
frustration and 3) quantum fluctuations enhanced by low-dimensionality of 
the magnetic system. According to high temperature susceptibility measurements 
\cite{cvx_vx2_Niel_77}, the moment of V$^{2+}$ can be estimated as 3.96, 4.07 
and 4.07~$\mu_B$ for VCl$_2$, VBr$_2$ and VI$_2$ respectively, suggesting 
small covalency reduction. From the low-temperature neutron diffraction data 
an average moment in VCl$_2$ is $<S>/S=0.80\pm0.06$ \cite{vcl2_Kadowaki_87}, 
and in VBr$_2$ $<S>/S=0.83\pm0.04$ or $<S>/S=1.02\pm0.05$ depending upon 
magnetic structure assumed \cite{vbr2_Kadowaki_85}. The relatively small reduction 
of the magnetic moment in the two-dimensional systems in comparison with 
almost 50\% reduction in some ABX$_3$ one-dimensional systems (see next 
section) indicates dimensionality (quasi-two-dimensional rather than 
quasi-one-dimensional) and quantum fluctuations are the   major influences 
on the moment reduction in the stacked triangular lattice. Note, that 
in theory an average magnetic moment of two-dimensional magnetic system
is nonzero, while in one-dimensional system $<S>=0$.

\subsection{Heisenberg triangular antiferromagnet with Easy-Axis 
anisotropy}
\label{HEA}

In this section we describe materials with a Hamiltonian (equation \ref{H_H})
that has an exchange term and a single-ion anisotropy with $D<0$. 
This negative value of $D$ makes it energetically favourable for the 
moments to align perpendicular to the $ab$ plane. This breaks the isotropic 
symmetry of the Heisenberg Hamiltonian and leads to changes in the physics.

There are five materials with this Hamiltonian, all of composition ABX$_3$,
with space group $P6_3/mmc$ and a stacked triangular lattice of spins. 
These are CsNiCl$_3$, RbNiCl$_3$,
CsNiBr$_3$, RbNiBr$_3$ and CsMnI$_3$.  All exhibit the quasi-one-dimensional
behaviour typical of this crystal structure with the value of the exchange
interaction, $J$, along $c$ more than an order of magnitude greater than 
the intra-planar exchange interaction, $J^{\prime}$, and with the magnitude 
of $D$ of the 
same order as that of $J^{\prime}$. Experimental values of $J$, $J^{\prime}$ 
and $D$ were given in table \ref{Heis_parameters}. In all cases the exchange 
interaction is antiferromagnetic. In some cases there are major discrepancies 
between experimental results; we will look at this later in this section.

All five materials have a magnetic phase diagram as shown in 
figure~4. At low temperatures and magnetic fields the magnetic structure is 
observed to be the
triangular structure with the $c$ axis in the plane of the triangle. The
anisotropy favours alignment of spins perpendicular to the basal plane
resulting in structures where the angle $\theta$ in the figure is less than 
60$^{\circ}$. Each $ab$ plane has a net moment perpendicular to it which 
can be observed by neutron diffraction. However the planes are stacked
antiferromagnetically because of the large antiferromagnet exchange
interaction, $J$, along the chains. This results in the whole crystal 
being an antiferromagnet. For a classical system
\begin{equation}
cos \hspace{1mm} \theta = \frac{1}{2(1+\frac{D}{6J'})} \hspace{5mm},
\hspace{50mm} {\mid}D{\mid} < 3J' \hspace{3mm}and\hspace{3mm}D < 0
\label{theta}
\end{equation}
At ${\mid}D{\mid}=3J'$ the triangular structure collapses into a collinear
structure with two spins along $+c$ and one spin along $-c$. The anisotropy 
is not large enough for this to happen in any of the materials considered 
in this section. Chapter~\ref{Ising} describes collapsed cases. 

At a temperature $T_{N2}$ there is a critical phase transition to the colinear
structure as shown in figure~4. 
Thus for classical spins, as ${\mid}D{\mid}$ increases to 3$J'$, $T_{N2}$ decreases 
to zero. At a 
higher temperature, $T_{N1} > T_{N2}$, there is a second critical phase
transition, this time to the paramagnetic state. The difference
$(T_{N1}-T_{N2})/T_{N1}$ is a measure of the relative strength
of the anisotropy, $D$, and the exchange, $J'$. The colinear structure has
energy independent of applied field, while for all the competing structures 
the energy decreases as the field increases. Thus it becomes less stable as 
the field increases.

As a field is applied along the $c$ axis at low temperature the angle $\theta$ 
changes in different directions for successive planes; in the plane shown in
figure~4
 the angle $\theta$ increases on application of a field along $Z$.
At a field $H_c$ there is a first-order phase transition to the canted 
structure shown in figure~4. 
In this structure the $XY$ component of the moments form an equilateral
triangle, while the $Z$ components are aligned ferromagnetically along the 
field direction. For a classical system with $J$ large compared with $J'$ or
${\mid}D{\mid}$, Tanaka {\it et al.} \cite{abx_Tanaka_88} show that
\begin{equation}
(g \mu_B H_c)^2 = 16JDS^2
\label{Hc}
\end{equation}
at $T=0$. It should be noted that the expression given by Tanaka {\it et al.}
differs from that given above by a factor of 2 because they define $J$ to 
be half of our $J$.

There is a multicritical point at $T_M$, $H_M$ where all four phases meet.
Alternatively, this point can be described as the intersection of three lines
of critical phase transitions and one line of first-order phase transitions.
The experimentally-determined parameters of the phase diagram, $T_{N1}$, 
$T_{N2}$, $T_M$, $H_c$, $H_M$ and $\theta$ are given in table~\ref{HEA_tab}.
There is reasonable agreement between different experimental measurements 
here, in contrast to the descrepancies noted between some of the
measurements of the parameters of the Hamiltonian (table~\ref{Heis_parameters}).

Now that we have given an overview of the properties of materials with
easy-axis anisotropy, we go back to examining the experimental values of the
parameters in the Hamiltonian, and to checking consistency with phase-diagram
parameters.  One reason for inconsistency is that it is believed that the
neutron scattering values of -13.0~GHz for CsNiCl$_3$ and -0.5 for CsMnI$_3$
are  in error because of incorrect branch assignments. A second problem is that
the resonance experiments actually measure $H_c$ and then derive $D$ from
equation \ref{Hc} and known values of $J$. Unfortunately the equation may not
hold for $S=1$. The difficulty  arises from the need to incorporate quantum 
fluctuations into the theory. A well-established technique for taking these 
into account involves an expansion in powers of $1/(2S)$ \cite{Ogu_60,Lin_76}.  
Zhitomirsky and Zaliznyak \cite{Zhi_96} show that inclusion of this term in 
equation \ref{Hc} yields a negative value for $H_c^2$. Although this is
non-physical, the result serves notice that equation \ref{Hc} is not good for
values of $S$ as small as 1. Furthermore the first term in the quantum
correction for $S=1$ renormalizes $J$ by 18$\%$ and $D$ by -50$\%$. 
Most experiments will measure the renormalized values, not the bare values. 
A further difficulty with the nickel compounds is that because of their
quasi-one-dimensional properties, there may be vestiges of the Haldane 
effect in their low-temperature properties. Of course a full quantum 
treatment will take this into account, but it may not be apparent in the 
first term of a $1/(2S)$ expansion. Several neutron scattering experiments 
claim to see Haldane-gap effects in the three-dimensional ordered phase
\cite{cnc_Morra_88,cnc_Tun_90,rnc_Tun_91,cmi_cnc_Enderle_93,cmi_cnc_Enderle_94}
 and conventional spin-wave theory (i.e. spin-wave theory without quantum
corrections) does not fit the measured dispersion relations. Without further
theoretical guidance there must be some doubts about the reported values of $J$,
$J'$ and $D$ for the nickel compounds even in cases where the experimental 
data is irreproachable.  Affleck\cite{Affleck_89} has argued Haldane correlations 
will still be effective in the three-dimensionally ordered phase. His treatment 
shows much better qualitative agreement with experiment than linear spin-wave 
theory  and provides a theoretical underpinning for the claims that the experiments 
constitute solid confirmation of  the existence of the Haldane gap. In large fields, 
$H>T_c$ and $T_M$, the quantum fluctuations are 
reduced!\cite{t_Zaliznyak_92,cnc_Katori_95} and the interpretational difficulties 
are less severe.

In CsMnI$_3$ the quantum corrections are smaller; in leading order by a factor
of 2.5 so that they are of the same order as the experimental uncertainties.
The spin wave dispersion relations follow conventional spin wave theory 
\cite{cmi_Inami_94,cmi_Tun_94}. The values of $J$, $J'$ and $D$ quoted in 
the table are the directly measured values without application of any 
quantum corrections.

It is interesting to consider the effects of increasing departures of the
Hamiltonian from the Heisenberg form in easy-axis systems. This is equivalent
to increasing $|D|$ while keeping the exchange parameters $J$ and $J'$
fixed. Even small values of $|D|$ break the symmetry by aligning the
moments triangles in a plane that contains the $z$ axis and splitting the
critical point into two critical points at $T_{N1}$ and $T_{N2}$. $T_{N2}$
decreases until it becomes zero at $-D=3J'$. $T_{N1}$ and $T_M$ increase as
$|D|$ increases, as can be seen for instance by comparing CsNiCl$_3$
with CsNiBr$_3$  where $J$ and $J'$ are similar but the values of $D$ differ 
by an order of magnitude. $H_c$ increases as $|D|^{\frac{1}{2}}$ and experiment 
seems to indicate that $H_M \simeq 1.1 H_c$.

We now discuss the phase diagram and critical properties of the easy-axis
magnets. The theory is constructed assuming that the parameters of the
system are such as to give a phase diagram of the type shown in
figure~4
 with the magnetic structures that we have described. 
Thus antiferromagnetic Heisenberg interactions $J$ and $J'$ are taken on 
a  stacked triangular lattice with easy-axis anisotropy $-D<3J'$. Then, 
building on the ideas of chiral universality classes described earlier, 
theory makes a number of predictions based on Landau-type theories, 
scaling and renormalization group calculations. 

1. There is a multicritical point connecting a line of first-order phase
transitions and three lines of critical phase transitions \cite{t_Plumer_88}.

2. All three lines of critical points come in tangentially to the first order
spin-flop line \cite{t3_Kawamura_90}.

3. The lines of phase transitions through $T_{N1}$ and $T_{N2}$ both follow 
the $XY$ universality class \cite{t3_Kawamura_90}.

4.The phase transition from the paramagnetic state to the spin flop state
follows the chiral $XY$ universality class \cite{t3_Kawamura_90}.

5. At the multicritical point the phase transition should follow the chiral
Heisenberg universality class \cite{t3_Kawamura_90}.

6. The transition region around the multicritical point should not be large 
\cite{t_Kawamura_93}.

The first of these predictions is obeyed by all the materials studied in this
section. The experimental evidence from the highest accuracy phase diagrams in 
CsNiCl$_3$
\cite{cnc_Poirier_90,cnc_Trudeau_93}, CsNiBr$_3$ \cite{cmi_cnb_Katori_93} and
CsMnI$_3$ \cite{cmi_cnb_Katori_93,cmi_cnc_Enderle_94} suggests that the second 
prediction is also good. The experiments all
clearly show that the two lines of phase transitions to paramagnetism come
in parallel to the line of first-order phase transitions and give clear
indications that there is curvature in the line of phase transitions through 
$T_{N2}$ close to the multicritical point such as to make the predicted effect
likely.

Table \ref{hea_crit} lists the experimental determinations of the critical
exponents $\beta$, $\nu$, $\gamma$ and $\alpha$ and compares the results with
 predictions for various universality classes. As well as the critical
exponents, the table also compares experiment and theory for the amplitude
ratio, above and below $T_N$, of the specific heat divergence.

Inspection of the table shows that there is generally good agreement between
experiment and theory for the specific heat data, but that the agreement with
the neutron data at $H = 0$ for $\beta$, $\gamma$ and $\nu$ is not good. Recent 
neutron measurements \cite{cnc_Enderle_97} for $\beta$ at the multicritical 
point and for the spin-flop-to-paramagnetic phase transition agree well with 
theoretical predictions.

We conclude that the theoretical predictions 1,2,4 and 5 above are confirmed by 
experiment. Prediction 6 has not been tested. Experiment does not bear out  
prediction 3 that the critical phase transitions at $T_{N1}$ and $T_{N2}$ follow 
the $XY$ universality class.  The measured
indices do not fall within any standard universality class, but they do seem to
be the same at the two transitions as predicted. The neutron data could be
described numerically by  chiral Heisenberg critical exponents, suggesting that
the crossover from the multicritical point takes place slowly, but the specific
heat data  and theoretical prediction number 6 both argue against this
possibility.

\subsection{Heisenberg triangular antiferromagnet with Easy-Plane type anisotropy}
\label{HEP}

The presence of an anisotropy term in the Hamiltonian with $D>0$ favours 
the confinement of spins to the $XY$ plane. The ordered structure in the 
absence of a field is an equilateral triangle of spins with two 
chirally-degenerate states as was shown in figure 1.
 For any atom with spin aligned 
along its local $z$ direction, the local $xy$ degeneracy of the Heisenberg 
Hamiltonian is broken since a rotation of spins in the plane costs no 
anisotropy energy while a rotation out of the plane does cost anisotropy 
energy. This splits a degeneracy in the spin wave excitations, with one 
acoustic branch having zero energy at magnetic reciprocal lattice points 
while the other exhibits a gap.

Near the critical point the fluctuations will tend to diverge in the $XY$
plane, but not in the $Z$ direction, so that the critical exponents will 
correspond to the chiral $XY$, or $Z_{2}{\bf \times}S_{1}$, model. If $D$ is 
small there should be a crossover to Heisenberg exponents further away 
from the critical point.  This effect has not been seen to date since only 
one easy-plane material, CsMnBr$_3$, has had its critical exponents 
measured extensively, and in this material $D$ is not small compared with $J'$.

The behaviour of easy-plane triangular antiferromagnets in a small field 
perpendicular to $Z$ involves a competition between the in-plane exchange 
energy $J'$ and the anisotropy energy $D$. The $J'$ term is of lowest 
energy when the moments are aligned in the 120$^\circ$ structure 
perpendicularly to the field with a canting of the moments towards the 
field direction, while 
the anisotropy term favours 120$^\circ$ structure aligned in the $XY$ 
plane. What happens as the field is increased depends on the relative 
magnitudes of $D$ and $J'$ \cite{t_Chubukov_88}. At low temperatures, if 
$D<3J'$, a field in the $XY$ plane larger than $H_s$ will flip the plane 
of the spin triangle  so that it is perpendicular to  the 
field. This costs anisotropy energy, has virtually no cost in exchange 
energy and 
gains energy from a canting of the spins along the field direction. The 
spin-flop phase transition will be of first order at
\begin{equation}
  g \mu_B H_s = 4S\sqrt{JD}          
\end{equation}
If the anisotropy energy is larger, $D>3J'$, the 
ground state above a critical field $H_c$ is one where the spins remain in 
the plane, but the triangular structure collapses to a colinear structure 
with two spins in one direction in the plane normal to the field direction 
and one spin in the opposite direction. This structure costs exchange 
energy $J'$, has no cost in anisotropy energy, and gains energy by a 
canting of the spins towards the field direction. The phase transition 
will be continuous with
\begin{equation}
  g \mu_B H_c = 4S\sqrt{3JJ'}
\label{Chub_Hc}
\end{equation}
It is clear that of these two cases, it is the latter with $D>3J'$ that 
corresponds the more closely to the chiral $XY$ model.

Table \ref{HEP_table} lists the easy-plane triangular 
antiferromagnets. All have the ABX$_3$ structure with strong exchange 
interactions $J$ and relatively weak in-plane interactions $J'$. In the 
table we give the type of antiferromagnet, the space group, the N\'{e}el 
temperature, the critical magnetic field and the aligned magnetic moment 
at low temperatures.

In the rest of this section we discuss separately the three cases: small $D$,
large $D$ and those where the crystal structure is distorted.

\subsubsection{The case of small anisotropy $D<3J'$}

There are four materials which are known to be in this category, CsVCl$_3$,
CsVBr$_3$, CsVI$_3$ and RbVCl$_3$. The crystal and magnetic structures were
established by Zandbergen \cite{abi_Zandbergen_81} and by Hauser
{\it et al.} \cite{avx_Hauser_85}. Low field susceptibility measurements by
Niel {\it et al.} \cite{avx_Niel_77} on powder samples in the
quasi-one-dimensional region ($T \gg T_N$) established the intrachain exchange
parameter $J$ for the cesium compounds. Feile {\it et al.}
\cite{cvc_cvb_cbi_Feile_84} have measured the spin wave dispersion relations in
CsVCl$_3$, CsVBr$_3$ and CsVI$_3$ by neutron inelastic scattering. The
measurements were confined to the ($\xi \xi 1$) direction of reciprocal space at
low temperatures. The results fitted reasonably with the predictions of linear
spin wave theory given by Kadowaki {\it et al.}\cite{cvc_Kadowaki_83} for
nearest neighbour interactions and single-ion anisotropy. The restriction of
the data to the plane results in the fitting only giving values for $JJ'$
and $JD$ and a value of $J$ was taken from the paramagnetic magnetization
measurements of Niel {\it et al.} in order to determine the values of 
$J^\prime$ and $D$ given in Table~\ref{HEP_table}. No
zero-point-motion correction was made for the reduced moment and
one-dimensionality so the real values of $|JJ'|$ and $|JD|$ are lower
than  those quoted. On average the observed ordered moment on the vanadium atom, as
taken from Table~\ref{HEP_table}, is 1.9~$\mu_B$. For $S = 3/2$ with $g = 2$ the maximum
moment is 3~$\mu_B$, so the reduction is 63\%\ and the correction to linear
spin-wave theory is appreciable.

Little is known experimentally of the phase diagrams or critical properties of
these materials. There has been no investigation of the nature of the phase 
transition in zero field.  As pointed out earlier, this is an unresolved area 
in the theory, with predictions of chiral XY critical exponents, tricritical 
exponents and a weak first-order phase transition in the literature. It 
would be interesting to discover what experiment has to say.

The phase diagram for a field $H$ applied in the $XY$ plane is expected to be as
shown in figure~\ref{HEP_smallD_PhD}. Phase~$I$ is the plane triangular 
antiferromagnetic structure, phase~$II$ is the colinear structure, phase~$III$ is 
the spin-flopped triangular structure and phase~$P$ is paramagnetic.
Molecular field calculations by Plumer {\it et al.} \cite{t_Plumer_89} predict
this phase diagram or similar ones differing only by the presence of a narrow
region of an extra phase near the first order transition line. The prediction
was actually for the $XXZ$ Hamiltonian which is slightly different from our
Hamiltonian~(Eq.\ref{H_H}), but the differences are not expected to affect the overall
pattern significantly. At $H=0$ and $T=T_N$ there is a meeting of four 
phases  since at negative $H$ there is a similar phase to phase~$II$ with
the canting in the opposite direction; thus it is a tetracritical point. The
tetracritical nature of the transition shows that it does not belong in the
same universality class as the $XY$ model.
 
Only one critical exponent, $\phi$, has been determined in these materials. Near
the tetracritical point the two phase boundaries are predicted to behave as
\cite{t_Kerszberg_78}
\begin{equation}
H^2 \simeq |T_{Ni}(H^2) - T_N|^{\phi_i},
\end{equation}
where $i$ has one of two values depending on whether $T_{Ni}$ refers to the
phase boundary~$II$ to paramagnetic or $I$ to $II$. Actual determination of
these exponents is quite sensitive to the value chosen for $T_N$ and to the
range of data over which the fit is made \cite{Gaulin_94}. The sum of the two
exponents $\phi$ is much less sensitive to the value of $T_N$ than is the
individual values of $\phi$.

In CsVBr$_3$, Tanaka {\it et al.}~\cite{cvb_rvb_Tanaka_94} have determined the
critical exponents $\phi$ in a field perpendicular to the $c$ direction from
susceptibility data. They find $\phi_{P-II}=0.78(6)$ and $\phi_{I-II}=0.79(6)$. 
Scaling and renormalization group theory
\cite{Kawamura_88,t_Kawamura_89,t3_Kawamura_90} gives 
$\phi_{P-II}=\phi_{I-II}=\phi$, with 
$1<\phi<\gamma$; for the $XY$ chiral model $\gamma$ is expected 
to be only slightly greater than one, probably
about 1.1. Thus the experimental values of $\phi$ are significantly lower than
predicted by theory. The discrepancy cannot lie in the sensitivity of the fits
to the value of $T_N$, since the sum of the two individual values of $\phi$ is
just as far from the prediction as is the individual values.

\subsubsection{The case of large anisotropy $D>3J'$}

This is the unusual and probably the most interesting case. Although there 
is only one example of undistorted triangular easy-plane antiferromagnet 
which satisfies this condition, CsMnBr$_3$, it has attracted a lot of both 
theoretical \cite{cmb_Oyedele_78,t_Chubukov_88,cmb_Mason_90b,t_Abarzhi_93} 
and experimental attention through neutron scattering 
\cite{cmb_Eibshutz_72,cmb_Collins_84,cmb_Gaulin_87,cmb_Mason_89,cmb_Gaulin_89,cmb_Mason_90a,cmb_Mason92}, magnetization 
\cite{cmb_Goto_90,cmb_Kotyuzhanskii_91,cmb_Abarzhi_92}, 
ESR \cite{cmb_cnc_Zaliznyak_90} and specific heat measurements 
\cite{rev_Weber_95}. As was pointed out in the introduction to this chapter, if $D>3J$' 
the application of a field perpendicular to $c$ no longer produces the spin-flop phase
shown in fig.~\ref{HEP_smallD_PhD}. Instead there is a critical phase transition to an
almost colinear structure at a critical field which is given by equation~\ref{Chub_Hc} 
at $T=0$. The phase diagram is shown in fig 6. 
 At low
temperatures and fields the magnetic structure is the 120$^{\circ}$ stacked
triangular structure  with the anisotropy confining the moments to within the
$ab$ plane.  The critical point at $T=T_N$ and $H=0$ is a tetracritical point just
as for the case $D<3J$' described in the previous section. In the high-field phase
two of the three moments on the triangular lattice become parallel with the
third in the antiparallel direction. The moment directions are in the basal
plane almost perpendicular to the applied field, but with a small canting
towards the field direction proportional to $H/J$.  In this phase there is a
softening of one of the exchange branches found by ESR
\cite{cmb_cnc_Zaliznyak_90}.

The nature of the phase diagram was discovered through neutron-scattering 
measurements of the temperature dependence of the intensity of magnetic Bragg 
peaks in different fields as shown in fig.~\ref{CsMnBr3_int}
\cite{cmb_Gaulin_89}. These results were later confirmed by measurements of 
the field and temperature dependence of the magnetization as shown in 
fig.~\ref{CsMnBr3_M} a) and b).
The magnetization process is described satisfactorily in terms of linear
spin-wave theory except for two features: (i) the measured torques are
significantly smaller than predicted and (ii) in high fields, H$>$H$_c$, there
is a considerable anisotropy between the magnetization when the field is
applied along and perpendicular to the $c$ axis. It has been suggested by Abarzhi
{\it et al.} \cite{cmb_Abarzhi_92} that quantum fluctuations are responsible 
for both these effects. This claim has bee supported  by subsequent work
\cite{rmb_Abanov_94,t_Zaliznyak_92} but recently has beem disputed by Santini
 {\it et al.}~\cite{t_Santini_96}, at least  for CsMnBr$_3$ and RbMnBr$_3$, where 
 they claimed that
 the anisotropy is already present at the classical level, provided thermal
 fluctuations are taken into account. Santini's work  is based on a Hamiltonian
 similar to (\ref{H_H}) with  the $J'$ term replaced by 
 a dipole-dipole interaction.

A peculiar feature of the phase transition from a triangular to a colinear 
structure is that it survives even if the field direction deviates 
significantly from the basal plane. The value of the critical field follows the equation
\begin{equation}
H_c^2 (\varphi) = H_c^2 \frac {d-1}{d \cos^2(\varphi)-1}
\end{equation}
where $d=D/3J'$ and $\varphi$ is the angle between the magnetic field and the
basal plane, rather than simple field projection on the spin's plane,
$H_c(\varphi)=H_c\cos{\varphi}$~\cite{cmb_Abarzhi_92}.

Kawamura \cite{t_Kawamura_87,t_Kawamura_89} predicted that the easy-plane
materials would not follow the $XY$ universality class in their critical
properties in zero field because of the extra chiral degeneracy of the order
parameter.  Instead a new universality class, known as the chiral $XY$ class,
would apply. The discovery that the critical point is tetracritical confirms in a
simple way that regular $XY$ universality does not apply.  The 
most recent Monte-Carlo work, done for the case of ferromagnetic 
interactions along the c direction, favours a weak first-order phase 
transition \cite{Plumer_94, Boubcheur_96,Plumer_96}, with effective critical exponents 
that are reasonably in agreement with  those given by Kawamura for the chiral XY model, 
less than three errors apart in all cases.

All the known undistorted easy-plane triangular materials are reported 
experimentally to show critical phase-transitions in zero field in 
agreement with Kawamura's predictions. Of course it is never possible 
experimentally to rule out a first-order phase transition if it is 
sufficiently weak, but critical behaviour is observed to persist at least 
down to reduced temperatures of 2$\times$10$^{-3}$ \cite{rev_Weber_95}. 
 The measured values of the critical indices are listed in 
table~\ref{critind} and compared with the predictions of four possible models: chiral
$XY$, chiral Heisenberg, regular $XY$ and tricritical in mean-field theory. The
exponent $\overline{\phi}$ is the mean of  $\phi_{P-II}$ and $\phi_{I-II}$.
We use this parameter because Gaulin \cite{Gaulin_94} has shown that while 
experimental determinations
 of  $\phi_{P-II}$ and $\phi_{I-II}$ are highly sensitive to the value of $T_N$, the 
mean of the two is much less sensitive, and hence is determined more reliably
in experiments. The exponent $z$ is the dynamic critical exponent measured from
the temperature dependence of the energy of long-wavelength spin waves 
\cite{rev_Collins_89} near $T_N$.

The experimental results fit both the chiral $XY$  and the mean-field tricritical 
models
satisfactorily, while they do not fit the other two models. It is not
understood why theory and experiment agree so well for the easy-plane case, but
less satisfactorily for Heisenberg and easy-axis materials.

\subsubsection{Distorted crystal structures}
\label{Distort}

It is not uncommon for the magnetic ABX$_3$ compounds to experience a 
crystal distortion. A different types of crystal structure distortions 
have been found at sufficiently low temperature in KNiCl$_3$, RbMnBr$_3$, 
RbVBr$_3$, RbVI$_3$, RbTiI$_3$ and RbFeBr$_3$. Generally lattice 
deformations due to a structural phase transition lead to some 
modifications of magnetic interactions (the crystal distortions break the 
symmetry and so change the exchange interaction between neighbouring 
in-plane magnetic ions) and, consequently, to a partial lifting of 
frustration on a stacked triangular lattice.  Study of such partially 
frustrated systems is of a fundamental interest, because they do not 
simply correspond to an intermediate case between unfrustrated and 
frustrated magnets but show novel physical phenomena absent in the two 
limiting cases.

Phase transitions to lattices of lower symmetry with decreasing 
temperature are characterized usually by displacements of chains of 
magnetic atoms as a whole without deformation so that the intrachain 
distance between spins remains unchanged.  The typical structural 
transition to the lattice of $P6_3cm$ space group is accompanied by the 
shift of one from the three adjacent chains upward along the $c$-axis 
while the two others shift downward keeping the crystal center of mass 
undisplaced. This primary distortion is shown in figure \ref{knicl_dist}.

The crystal unit cell in the basal plane is enlarged to become $\sqrt{3}a 
\times \sqrt{3}a$ (Fig.\ref{knicl_dist}), preserving the hexagonal 
symmetry. Because chains are shifted usually on a small distance from the 
basal plane ($\sim 0.5$ \AA\ in RbFeBr$_3$ \cite{rfb_Eibschutz_73}), 
magnetic properties may be considered by placing spins on the same stacked 
triangular lattice and changing interactions in the Hamiltonian 
(\ref{H_H}) in accordance with a reduced symmetry of the crystal 
structure. The distortion has little effect on the exchange $J$ along the $z$ 
direction 
but in the XY plane the interaction $J'$ is split into two different 
interactions, $J'_{AB}=J'$ and $J'_{AA}=J'_1$ as shown in figure \ref{mag_dist}a. 
The $120^{\circ}$ 
triangular structure corresponds to $J'=J'_1$ and small departures 
from this condition give triangular structures with angles not equal to 
120$^\circ$. Mean-field investigations have been carried out to determine 
the full phase diagram as the ratio of $J'$ to $J'_1$ is changed 
\cite{t_Kawamura_90,t_Zhang_93}. 

The Hamiltonian of this ``centered honeycomb model" (in terms of 
Zhang {\it et al}. \cite{t_Zhang_93}) is obtained by the evident 
replacement of the second term on the r.h.s.\ of Eq.~(\ref{H_H}): 
\begin{equation}
J'\sum_{i,j}^{\text{A-A}}{\bf S}_i{\bf S}_j + 
J'_1\sum_{k,l}^{\text{A-B}}
{\bf S}_k{\bf S}_l \ . 
\label{honey}
\end{equation}

Without field the spin ordering occurs in two steps with additional 
intermediate collinear phase between $T_{N1}$ and $T_{N2}$ which is either 
ferromagnetic or partially disordered \cite{t_Plumer_91}. The splitting of
$T_N$ was clearly seen in RbVBr$_3$ \cite{cvb_rvb_Tanaka_94}.
The behaviour of the system in the applied magnetic field depends on the 
relative strength of exchange constants $J'$ and $J'_1$. Note, that due to
superexchange character of the interchain interaction, it depends in a
complicated way from the interatomic distances and bond angles. The critical 
exponent $\beta$ at T$_{N1}$ is  observed \cite{cvc_Kakurai_9?} to be 0.32(1), 
while theory predicts that this phase transition should follow the XY model 
with $\beta$ = 0.35.

If $J'<J'_1$, that is coupling between in-plane spins (A$_1$--A$_2$ in
fig.~\ref{mag_dist}a) is stronger than coupling between in-plane and
out-of-plane spins (A--B), then the presence of the distortion does not 
change nature of the phase transition to a colinear phase. Critical field
is given in this case by the formula \cite{knc_Tanaka_89}:
\begin{equation}
(g \mu_B H_c)^2 = 48 J (2J'-J'_1) S^2,          
\end{equation}
which is very similar to a formula~(\ref{Chub_Hc}) when $J'$ and $J'_1$
are close to each other.

If $J'<J'_1$, then at low magnetic field spin of the out-of-plane atom (marked
B in fig.~\ref{mag_dist}a) is aligned parallel to the field. Such a
configuration is energetically unfavourable at higher fields, therefore
an additional phase transition occurs at $H^\star < H_C$, when sublattice 
B starts to deviate
from the field direction \cite{rmb_knc_Zhitomirsky_95}. Finally, the transition
to a collinear phase occurs at   
\begin{equation}
(g \mu_B H_c)^2 = 48 J (\sqrt{J'^2+3J'} - J') S^2,          
\end{equation}
which is again very similar to a $H_c$ in undistorted triangular structure.

The above described theory was developed to explore the magnetic consequences
of the crystal phase transition  $P6_3/mmc \rightarrow P6_3cm$. Distortions of 
this type were found in the low temperature 
phase of RbFeBr$_3$ \cite{rfb_Eibschutz_73}, at the room temperatures in 
KNiCl$_3$ \cite{knc_Visser_80}, and in RbMnBr$_3$ \cite{rmb_Fink_82} and, 
probably, in RbVBr$_3$ \cite{rvb_Tanaka_94}. But, as it will be evident
from the discussion given below, in at least two compounds, RbMnBr$_3$ and
KNiCl$_3$, further crystal phase transitions just below room temperature
play an important role.

In KNiCl$_3$ dielectric anomalies indicating structural phase transitions 
are found  at 274~K, 285~K, 561~K and 762~K \cite{knc_Machida_94}.
A single crystal x-ray study on the {\it low-temperature} structure of 
KNiCl$_3$ \cite{knc_Petrenko_96b} shows clearly the existence of two crystal 
structure distortions, as originally found by neutron scattering
 \cite{knc_Petrenko_96a}.
One phase (denoted as phase {\it A}) is hexagonal and does not differ 
much from the room temperature structure, the other phase (phase {\it B}) is 
orthorhombic. In a phase {\it A} 
 the unit cell is  rotated through $90^\circ$ about the $c$ axis 
from the room temperature unit  cell and enlarged to $\sqrt{3}a$, 
$\sqrt{3}a$ and $c$; in a phase {\it B} the low temperature unit cell has 
sizes $2a/\sqrt{3}$, $a$ and $c$. The main feature of the phase 
{\it B} is a sinusoidal modulation of the ion chains in the basal plane: 
instead of the room temperature sequence 0-0-UP-0-0-UP-0-0, where ``0" means 
ion in the basal plane and ``UP" means ion slightly shifted above the basal 
plane along the $c$-axis, at low temperature the sequence is
0-UP-0-DN-0-UP-0-DN-0, where ``DN" means the ion is shifted below basal 
plane. Possible space groups are $Pca2_1$ and $Pbcm$ \cite{knc_Petrenko_96b}.

As a consequence of the existence of two different crystal
modifications, two different magnetic structures have been
observed with $T_N = 12.5$~K and $8.6$~K in phases {\it A} and {\it B}
respectively \cite{knc_Petrenko_96a}. Magnetization 
measurements~\cite{knc_Petrenko_95} and measurements of 
ESR~\cite{knc_Tanaka_89,rmb_knc_Zhitomirsky_95} showed that in phase {\it A}
magnetic structure is a distorted triangular with $J'<J'_1$. 
Magnetic structure of phase {\it B} we discusse below, for now we just note, 
that in both cases magnetic structure is commensurate.

In RbMnBr$_3$ the situation with crystal distortion is very similar: 
the neutron measurements of Heller {\it et al.} \cite{rmb_Heller_94} show a 
crystal which seems to contain both the {\it A} and {\it B} phases, while 
the measurements of Kato {\it et al.} \cite{rmb_Kato_93} show a crystal 
where only the {\it B} phase is present. Moreover, recent x-ray scattering 
measurements \cite{knc_Petrenko_96b} shows total identity of RnMnBr$_3$ and
KNiCl$_3$ phase {\it B} crystal structures. Such an identity of crystal
structure makes it hard to explain difference in magnetic structure. From 
a symmetry point of view {\it B}-phase corresponds to the ``row model" of 
Zhang {\it et al.} \cite{t_Zhang_93} shown on fig.~\ref{mag_dist}b. 
Anticipation of this model results in appearance of incommensurate magnetic 
structure \cite{t_Kawamura_90,t_Zhang_93}. 

In RbMnBr$_3$ the magnetic structure is indeed incommensurate at low magnetic 
fields and only if the magnetic field exceeds 3~T it became commensurate 
\cite{rmb_Heller_94}. Incommensurate-commensurate phase transition is 
accompanied by the hysteresis phenomena in magnetization \cite{rmb_Bazhan_93},
resonance power absorption \cite{rmb_Vitebskii_93} and magnetic Bragg-peaks
intensity \cite{rmb_Heller_94}. The overall $H-T$ phase diagram, which is
much complicated and includes two incommensurate phases, two commensurate 
phases and paramagnetic one, was successfully explained in terms of Landau 
theory using ``row model" \cite{rmb_Zhitomirsky_96}. 

In phase {\it B} of KNiCl$_3$ the magnetic structure is commensurate even 
in a zero magnetic field \cite{knc_Petrenko_96a} and identical to 
the high-field structure of RbMnBr$_3$. What causes the stabilization of 
the commensurate spin configuration in KNiCl$_3$ remains unknown.

\section{Ising antiferromagnet}
\label{Ising}

The triangular antiferromagnets of type ABX$_3$ with B a  cobalt atom have
properties that are like those of Ising antiferromagnets. The cobalt cation lies
in an octahedron of X anions with a slight trigonal distortion.  The strong
crystal field splits the lowest lying $^4$F configuration so that a $^4$T$_1$
state has the lowest energy. This corresponds to a Kramer's doublet which is
effectively an $S=\frac{1}{2}$ state with the moment lying either parallel or
antiparallel to the $c$ axis.  There is a mixing between this state and a state
of higher energy also with $^4$T$_1$ symmetry and the resultant exchange
Hamiltonian at low temperature can be described by the equation
\cite{ccb_Nagler_83} 
\large 
\begin{equation}
\hat{\cal H} = J\!\sum_{i,j}^{\text{chains}}[ S^z_i S^z_j + \epsilon (S^x_i
S^x_j + S^y_i S^y_j)] + J'\!\sum_{k,l}^{\text{planes}}S^z_k S^z_l - g \mu_B 
H\sum_i S^z_i 
\label{H_I}
\end{equation}  
\normalsize
\noindent
with $0 < \epsilon < 1$. In every case $\epsilon$ is small, about 0.1, so that
the first term in the Hamiltonian, which is of Ising type, is the dominant
term. As in all the ABX$_3$ triangular systems the magnitude of the inter-chain
exchange constant $J'$ is
small compared with the exchange constant $J$ along the chains. $H$ is an external
magnetic field applied along the $z$ direction.  A weak single-ion
exchange-mixing term can be neglected for the purposes of this article. The
parameter $\epsilon$ has a weak temperature dependence, decreasing as the
temperature is raised \cite{ccb_Nagler_83}.
 
There are four ABX$_3$ compounds that have the magnetic Hamiltonian given
above, CsCoCl$_3$, RbCoCl$_3$, CsCoBr$_3$ and RbCoBr$_3$. There has been
extensive work on the first three of these, and table \ref{Ising_par} lists the
experimentally-determined values of the parameters in the Hamiltonian for each
compound. We also list values of phase-transition temperatures in zero field, 
$T_N$, and of applied fields, $H_c$, at which there are phase transitions.

Frustration effects are more acute in Ising triangular antiferromagnets
than in XY or Heisenberg systems. In the latter cases the frustration can be
partially relieved by the formation of spin triangle as was shown in figure
\ref{geom_frust} b) and c), while this is not possible for the Ising
antiferromagnet where spins are confined to directions parallel and
antiparallel to $z$. In contrast to the unfrustrated case, Wannier
\cite{Wannier_50} showed that the two-dimensional nearest-neighbour triangular
antiferromagnet is disordered at all finite temperatures and has a critical
point at $T=0$. Again the frustration changes the underlying physics.

In the stacked triangular lattice there is no frustration for nearest-neighbour
interactions along the stacking direction, whether they are ferromagnetic or
antiferromagnetic. This lessening of the overall frustration allows long-range
ordering at low temperatures.  Figure~\ref{ising_spin_arr}
shows three examples of possible ordering in the basal plane where sites 
marked $"+"$ have $S^z=\frac{1}{2}$, sites marked $"-"$ have 
$S^z=-\frac{1}{2}$, sites marked $"0"$ have $S^z$ randomly distributed 
with $<S^z>=0$, and sites marked $"\frac{1}{2}"$ have $S^z$ randomly 
distributed with $<S^z>=\frac{1}{2}S=\frac{1}{4}$. In each case the unit cell 
is $\sqrt{3a}$ by $\sqrt{3a}$, where $a$ is the lattice constant of the
triangular lattice. The states shown
in a) and c) are not fully ordered, while that shown in b) is fully
ordered and ferrimagnetic. There is no net ferrimagnetism over the whole
crystal in this state however since the strong antiferromagnetic exchange along 
$c$ ensures that the magnetic moment in a given plane is cancelled by an equal
and opposite moment on the next succeeding plane.

The three ordered arrangements shown in figure~\ref{ising_spin_arr} 
have the same energy, as do many other configurations, so the ground state has a
high degree of degeneracy. The unit cell for the three states is the same and
their neutron diffraction patterns are similar.  In real materials rather small
effects may enable the degeneracy to be split in favour of one particular 
ground state. For instance small ferromagnetic next-nearest-neighbour in-plane
interactions will stabilize state b), while such small antiferromagnetic terms
will stabilize state a).
 
In the known Ising ABX$_3$ compounds neutron diffraction in fact indicates the
presence of long-range magnetic order. There are sharp peaks of magnetic origin 
in the neutron-diffraction pattern below a temperature $T_{N1}$. These peaks 
are located in reciprocal space at ($\frac{h}{3}$ $\frac{h}{3}$ $\ell$), 
where $h$ is an integer not divisible by three and $l$ is an odd integer. 
The condition $l$ odd implies antiferromagnetic stacking along the $z$ 
direction, since the
ABX$_3$ structure has $c$ spacing equal to twice the interplanar spacing.  This
stacking reflects the strong antiferromagnetic coupling along the $c$ axis.
The $\frac{h}{3}$ factors indicate that the unit cell in the XY plane is
$\sqrt{3a}$ by $\sqrt{3a}$.  This is not an unexpected result in view of the
discussion earlier regarding figure~\ref{ising_spin_arr}. 
In figures~\ref{FYELON} and \ref{FMEKATA}, taken from Yelon {\it et al.}
\cite{ccob_Yelon_75} and Mekata and Adachi \cite{ccoc_Mekata_78} respectively
the temperature dependence of some observed neutron-scattering Bragg peaks is
shown for CsCoBr$_3$ and for CsCoCl$_3$. These measurements show that the
temperature dependence of the ordering is not simple. Some rearrangement of
the ordered structure at temperatures below $T_{N1}$is taking place, without
changing the unit cell or broadening the magnetic Bragg peaks.  Similar results
have also been reported by Farkas {\it et al.} \cite{ccb_Farkas_91} in
CsCoBr$_3$ and by Yoshizawa and Hirakawa \cite{ccc_Yoshizawa_79} in CsCoCl$_3$. 
 Detailed analysis has shown that the high temperature ordered structure in both
materials is similar to that shown in figure~\ref{ising_spin_arr} a).  Neutron critical 
scattering has been observed around $(\frac{1}{3} \frac{1}{3} 1)$ at temperatures close 
to $T_{N1}$ in CsCoBr$_3$ 
\cite{Rogge_95}.

As is shown in the figures, a feature of the low-temperature neutron scattering
 from  CsCoBr$_3$ and
CsCoCl$_3$ is the presence of a (111) magnetic Bragg peak.  This peak has the
scattering from the three magnetic sites in each basal plane in phase, while
there is a phase reversal between successive planes. Thus its intensity
reflects the net magnetic moment in each plane.  At temperatures below $T_{N3}$
(12~K in CsCoBr$_3$ and 5.5~K in CsCoCl$_3$) each plane is ordered
ferrimagnetically and the ordering corresponds to that shown in 
figure~\ref{ising_spin_arr} b).  
Neutron critical scattering has been observed around
 (111) at temperature close to $T_{N3}$ in CsCoCl$_3$ \cite{ccc_Boucher_85}.  
The absence of magnetic intensity in the (001) Bragg peak indicates that the
moments are ordered in the c direction.  At temperatures between $T_{N3}$ and
$T_{N2}$ =13.5~K in CsCoCl$_3$ there is clearly some structural rearrangement
taking place.  This is less evident in CsCoBr$_3$, but the data does seem to
indicate some small reduction of the magnetic intensities as the temperature is
lowered from $T_{N2}$=16~K \cite{ccob_Yelon_75,ccb_Farkas_91} to $T_{N3}$.  
Exactly what
is happening in the intermediate region between $T_{N2}$ and $T_{N3}$ has not
been established by the experimental work to date.
In assigning values to $T_{N2}$ and $T_{N3}$ in CsCoBr$_3$ we have
reinterpreted the data of Yelon {\it et al.} to be on a consistent 
basis with the assignments by other authors
\cite{ccb_Farkas_91,ccoc_Mekata_78,ccc_Yoshizawa_79}.  

Two experimental features should be mentioned here. First in CsCoBr$_3$ 
Yelon~{\it et al.} 
\cite{ccob_Yelon_75} indicate that there is also a small component of 
ordered moment in the XY plane, but this has not been reported in 
the work on RbCoBr$_3$ or CsCoCl$_3$.  If confirmed, such ordering would arise
from the term involving the parameter $\epsilon$ in equation (\ref{H_I}) and
indicates departure from the Ising form. Second,
a perhaps surprising experimental result is that, although the specific heat has
a clear anomaly at $T_{N1}$, there is no feature corresponding to the
transitions at $T_{N2}$ and $T_{N3}$ \cite{ccob_Yelon_75,ccb_Wang_94}.

Theory has concentrated on the solution to the true Ising system on the basis
of the 
Hamiltonian given in equation (\ref{H_I}) with $\epsilon = 0$.  There has also been
work with next-nearest-neighbour interactions in the basal plane included, so
as to reduce the ground state degeneracy.  The ground state in the
Landau-Ginzburg-Wilson (LGW) model is case~c) of 
figure~\ref{ising_spin_arr}~\cite{Berker_84}. 
There is a transition at temperature $T_{N2}$ to a second
ordered state corresponding to case~a) of figure~\ref{ising_spin_arr}
where one sublattice has a zero mean value of $S^z$. Plumer, Caill\'{e} and
Hood \cite{Plumer_89} show that, if the LGW treatment is expanded to higher
orderin the spin density, the phase transition can split into two transitions
at $T_{N2}$ and $T_{N3}$, although the splitting, $(T_{N2}-T_{N3})/T_{N3}$
is small, of order 1\%.  The specific heat effects are of opposite sign and tend
to cancel, agreeing with the experimental findings.

As discussed earlier,  experiments  
\cite{rcb_rnc_Minkiewicz_71,ccc_Melamud,ccoc_Mekata_78}
at low temperatures indicates the presence of an ordered state that corresponds
 to case b) of figure~\ref{ising_spin_arr}
 in contradiction to the predictions of the LGW model. Kurata and Kawamura
\cite{Kurata_95} have
recently shown that an extension of mean-field theory to include correlation
effects in the XY plane can give the observed ferrimagnetic ground state.

Even after this difficulty is taken care of, however, the LGW treatment still
has shortcomings  as it does not give the correct 
low temperature state, which Coppersmith \cite{Coppersmith_85} has shown 
involves some disorder on every site.  The various experiments refered to
above report similar values of the ordered moment at low temperature with a 
mean of 3.2(2)~$\mu{_B}$.  There is no appreciable difference between the 
predicted maximum 
ordered moment of $g_{\parallel} \mu{_B} S$ \cite{ccoc_ccob_rcol_Hori_90} 
and the 
measured moment, so the amount of the disorder is not large. However recent 
NMR work of Kohmoto
{\it et al.} \cite{Kohmoto_95} shows direct evidence of the 
presence of disorder at low temperatures as predicted.  

Another approach to the solution of the frustrated Ising model is to use
Monte-Carlo methods.  Matsubara $\it{et}$ $\it{al.}$
\cite{Matsubara_83,Matsubara_84,Matsubara_87} have shown that this gives the
ferrimagnetic phase at low temperatures  and two other ordered phases at higher
temperatures, in good agreement with experiment.
The high-temperature ordered phase, between $T_{N1}$ and $T_{N2}$, is a
randomly modulated phase (RMP).  Although the long-range order persists on a
$\sqrt{3a}$ by $\sqrt{3a}$ cell, there is a random modulation of $S^z$ on all three
sites.  The order parameter is the magnetic structure factor, 
\large
\begin{equation}
f({\bf Q}) = \!\sum_{j} S^z_j e^{i{\bf R}_j.{\bf Q}},
\label{f(Q)}
\end{equation}
\normalsize
\noindent
with ${\bf Q}=(\frac{1}{3} \frac{1}{3} 1)$. The low-temperature structure,
with T$<T_{N3}$, also has $f({\bf Q})$ as the order parameter, but with
${\bf Q} =(001)$.  The transitions at $T_{N1}$ and $T_{N3}$ are critical phase
transitions, but the nature of the transition at $T_{N2}$ is not clear. 
Between $T_{N2}$ and $T_{N3}$ the structure is complex with the characteristics
of the RMP phase present. but with an anomalous temperature dependence of the
order parameter $f(\frac{1}{3} \frac{1}{3} 1)$. Neutron scattering is a direct
technique for characterizing the phase transitions at $T_{N1}$ and at $T_{N3}$
because it can measure the scattering around $(\frac{1}{3} \frac{1}{3} 1)$
and (001) directly.  It is an unanswered question whether there is a true
phase transition at $T_{N2}$,  though there clearly is a region just above
$T_{N3}$ where the magnetic order has unusual temperature dependence. 
There is qualitative agreement between the neutron-scattering data and the
Monte Carlo work both with regard to the temperature dependence of the magnetic
structure factors and  to the ratio of $T_{N1}$ to $T_{N2}$ and to $T_{N3}$. 
Further the Monte Carlo computations show no specific heat anomaly at $T_{N2}$
or at $T_{N3}$, in agreement with experiment.

The nature of the phase transition at $T_{N1}$ has received much attention,
both theoretically and experimentally. Berker {\it et al.}
\cite{Berker_84} used renormalisation group arguments to predict that the
transition is in the same universality class as the order-disorder phase
transition in the three-dimensional XY model.

Early Monte Carlo work on the phase transition was done by Matsubara and
Inawashiro \cite{Matsubara_87} and by Hienonen and Petschek \cite{Heinonen_89}
but the most accurate analysis comes from the work of Bunker 
{\it et al.}~\cite{Bunker_93} 
and Plumer and Mailhot \cite{PlumerM_95}.
There is controversy concerning the accuracy of some of these results, with
Plumer and Mailhot's work showing satisfactory agreement with the
three-dimensional XY model and Bunker {\it et al.} showing
significant discrepancies. Table~\ref{I_crit} lists the values obtained by 
these authors and a comparison with the XY model and with experiment. Apart from
those of Farkas {\it et al.}, the experimental results for $\beta$
cover the same range as the theoretical values, with a weak bias towards the
higher values.  The experimental value for $\alpha$ claims higher accuracy than
any of the theoretical values but again no definitive conclusions can be made.
The Monte Carlo simulations of Plumer {\it et al.} \cite{Plumer_93}
indicate that the critical region at $T_{N1}$ is smaller than usual due to the
proximity of another ordered phase that is would be stabilized by
next-nearest-neighbour interactions of order 10\% of $J^{\prime}$.  
The recent neutron scattering work of Rogge \cite{Rogge_95} on CsCoBr$_3$ shows 
results incompatible with a normal critical
phase transition, in that the critical fluctuations cannot be described in
terms of a model with a single length scale.  This whole situation is  
unclear and more work is needed to resolve it.

Because of the Ising nature of the Hamiltonian and the quasi-one-dimensional
nature of the magnetic interactions, the excitations in these compounds are
predominantly of the soliton type.  Solitons have been observed in both
CsCoCl$_3$ \cite{ccb_Nagler_83,ccc_Boucher_85} and CsCoBr$_3$
\cite{ccb_Nagler_83,ccob_Nagler_83,ccb_Buyers_86} and the results used to
determine the parameters $J$ and $\epsilon$ in the Hamiltonian (equation
(\ref{H_I}) ).

Boucher {\it et al.} \cite{ccc_Boucher_85} show that soliton
excitations are present at temperatures down to $T_{N3}$.  The neutron
scattering data near the phase transition at $T_{N1}$ has been interpreted 
in terms of a soliton condensation on to one sublattice to give a magnetic
structure of the type shown in figure~\ref{ising_spin_arr} b)
with the solitons on the chains corresponding to the sites with zero mean
moment \cite{ccb_Tun_92,Gaulin_94}.  This result is not in accord with the 
idea of a randomly modulated
phase \cite{Matsubara_87} from Monte Carlo simulations.  Since all the Monte
Carlo work has been carried out for much-less-one-dimensional Hamiltonians
(smaller values of $J/J^{\prime}$) than is found in actual ABX$_3$ compounds,
and since the soliton ideas are products of the one dimensionality, the
soliton measurements \cite{ccc_Boucher_85,ccb_Tun_92} raise questions about
whether the Monte Carlo simulations map on to the real materials, particularly
as the excitations are fundamentally different.

We conclude this section with a discussion of the effect of a magnetic field on
triangular Ising antiferromagnets.  Consider first a one-dimensional Ising
antiferromagnet with nearest neighbour interactions. The Hamiltonian is
\large 
\begin{equation}
\hat{\cal H} = J\!\sum_{i,j} S^z_i S^z_j  - g \mu_B H\sum_i S^z_i 
\label{H_1D}
\end{equation}  
\normalsize
\noindent
Yang and Yang \cite{Yang_66} give an exact solution of this Hamiltonian, but
here we just treat the basic ideas.  At low temperature and H=0 the system will
form antiferromagnetic chains with a few solitons breaking the long-range
order.  The field will have only small effects until it reaches a critical
value $H_c$ where it can break antiferromagnetic bonds without cost in energy.
\large
\begin{equation}
g_{\parallel}\mu{_B}H_c = 2|J|
\end{equation}
\normalsize
\noindent
At field $H_c$ there should be a phase transition from antiferromagnetism to
ferromagnetism.  This transition is observed in the ABX$_3$ Ising compounds,
though the field $H_c$ is large ($\sim 40$~T) because the exchange $J$ along the
chains is large.  

In the ferrimagnetic-plane low-temperature structure (figure~\ref{ising_spin_arr} b) ) 
the small interchain exchange $J'$ results in there being two critical
fields, $H_{c1}$ and $H_{c2}$.  At field $H_{c1}=H_c$ one of the chains  
marked ``+'' in the figure becomes ferromagnetic without cost in interchain
exchange energy.  The magnetisation per cobalt atom is
$g_\parallel \mu_{B} S/6$.  Then at a higher field $H_{c2}$ the other two
chains become ferromagnetic, with
\large
\begin{equation}
H_{c2} = H_{c1} + 6|J'|/(g_{\parallel}\mu_{B})
\end{equation}
\normalsize
and the magnetization per cobalt atom is g$_{\parallel}$$\mu_{B}$S/2.

Figure~\ref{ising_H_c} shows the magnetization plotted against the applied 
field as observed in
CsCoCl$_3$ by Amaya {\it et al.} \cite{ccc_Amaya_90}.  The two steps in
the magnetization are not seen, but instead there is a rounding out between
$H_{c1}$ and $H_{c2}$ which is not expected from the simple arguments that we
have given. As the figure shows, this rounding becomes more pronounced at 
higher temperatures in the ordered phase. The critical fields, $H_{c1}$ and
$H_{c2}$ are usually identified with the two maxima in $dM/dH$, and it is these
values that are shown in table \ref{Ising_par}.

Table \ref{Ising_par} shows that the values of $J$ derived from $H_{c1}$ agree 
reasonably with those found by other experimental techniques, but the values of
$J'$ derived from $H_{c2}-H_{c1}$  seem to be significantly higher than
values from other experimental techniques.  They give $J/J'$ of order 10, which
is surprisingly small. Even the larger ratios found from neutron and Raman
scattering are appreciably smaller than is found in Heisenberg ABX$_3$
compounds (table \ref{Heis_parameters}), indicating that the cobalt compounds are 
less one
dimensional in magnetic properties.

\section{Singlet-ground-state magnets}
\label{SGS}
This chapter is devoted to the description of the magnetic properties of 
four compounds from the AFeX$_3$ family: CsFeCl$_3$, CsFeBr$_3$, 
RbFeCl$_3$ and RbFeBr$_3$. At room temperature they all have the same 
crystal structure with space group $P6_3/mmc$ and, as usual for all 
ABX$_3$ hexagonal compounds, at low temperature they exhibit 
quasi-one-dimensional magnetic behaviour. A characteristic property of 
these four crystals is the large value of the magnetic anisotropy in 
comparison with the exchange interaction.  In some cases this prevents the 
advent of long range magnetic ordering (LRO) even at zero temperature.  
There are other compounds (ND$_4$FeBr$_3$, ND$_4$FeCl$_3$, TlFeBr$_3$, 
TlFeCl$_3$ and CsFeI$_3$) that probably may be described as  
singlet-ground-state magnets \cite{nd4febr3_Harrison_91,afx_Visser_92}, 
but they have been investigated less thoroughly  and 
a comprehensive understanding of their physical properties has not yet been 
developed.

A free Fe$^{2+}$ ion in the AFeX$_3$ family has a $^5D$ ground state .
 A cubic crystal field splits this into an upper orbital doublet and a 
lower orbital triplet with an energy difference of order 1000~cm$^{-1}$.  Spin-orbit 
coupling, $\lambda '$, causes a further splitting of the triplet 
according to the effective total angular momentums ${\cal J}=1$, 2 and 3. 
The lowest state with ${\cal J}=1$ is split still further by a trigonal 
component of the crystal field, $\Delta '$, to produce a singlet ground 
state ($m_{\cal J}=0$) and an excited doublet ($m_{\cal J}=\pm 1$) as 
shown in Fig.\ref{sgs_split}. The Hamiltonian representing these 
splittings of the triplet state may be written as
\large
\begin{equation}
\hat{\cal H} = \Delta ' (L_{iz}'^2 - 2/3) + \lambda ' {\bf L'}_i {\bf S'}_i
\label{split_H}
\end{equation}
\normalsize
Since the energy separation between the ground state and the second 
excited state is of order of 100cm$^{-1}$ \cite{rfc_Eibschutz_75}, at low 
temperature only the first excited doublet is appreciably populated and  
the following effective spin Hamiltonian can be used to describe the magnetic 
properties of AFeX$_3$ compounds:
\large
\[
\hat{\cal H} =\!\sum_{i,j}^{\text{chains}}[J_\perp (S_i^x S_j^x +S_i^y S_j^y )
+ J_\parallel S_i^z S_j^z ] +
\!\sum_{k,l}^{\text{planes}}\![J'_\perp(S_k^x S_l^x+S_k^y S_l^y ) +
J'_\parallel S_k^z S_l^z ]  \]
\begin{equation}
+ D\!\sum_i(S_i^z)^2 - \mu_B \sum_i[g_\perp (S_i^x H_x + S_i^y H_y) + g_\parallel S_i^z H_z]
\label{SGS_H}
\end{equation}
\normalsize
where $S=1$ is fictitious spin and $D$, the value of which is positive, equals 
to the energy gap between the $m_{\cal J}=0$ and $m_{\cal J}=\pm 1$ 
states. However, some authors prefer to use the Heisenberg Hamiltonian 
(\ref{H_H}) to describe magnetic properties of the linear chains 
antiferromagneticaly coupled Fe$^{2+}$ ions in CsFeBr$_3$ and RbFeBr$_3$.

As $T\rightarrow 0$ in the absence of an external magnetic field there are 
two regimes separated by a phase transition.  

1. For $D<8\left| J\right|+12\left|J'\right|$ the system has a magnetic 
ground state with an easy-plane type of anisotropy.  This is the case for 
RbFeCl$_3$ and RbFeBr$_3$.  

2. For $D>8\left| J\right|+12\left|J'\right|$ the system has a singlet 
ground state and consequently does not order magnetically even at $T=0$. 
This is  the case for CsFeCl$_3$ and CsFeBr$_3$.  

The equality $D=8\left| J \right| + 12\left| J' \right|$ was derived as a 
condition at which the lowest excitation energy gap at the magnetic zone 
center becomes zero \cite{cfc_rfc_Yoshizawa_80}.

The application of an external magnetic field along $c$-axis on the SGS 
materials leads to a phase transition to an ordered state. This happens at a 
field $H_c$, when one of the excited doublet levels crosses  
the ground state singlet level, as it shown on Fig.\ref{sgs_h_split}. In 
CsFeCl$_3$ a commensurate 120$^\circ$ ordered structure appears after 
intermediate transitions through two 
incommensurate structures, while in CsFeBr$_3$ the  phase transition 
leads directly to commensurate order. If the external magnetic field is applied 
perpendicular to the $c$-axis, the singlet level remains below  
the excited levels at all fields so that  no LRO is expected. 

Before going to the detailed description of each compound we summarize 
some characteristics of each material in Table \ref{SGS_tab}. An attempt 
to analyze the correlation between the structural and magnetic parameters 
of the AFeX$_3$ family of compounds is made by Visser and Harrison in Ref.\cite{afx_Visser_88}. 

\subsection{The case of antiferromagnetic intrachain coupling}
\paragraph{CsFeBr$_3$}
In CsFeBr$_3$ and RbFeBr$_3$ all exchange interactions are antiferromagnetic.
Except for an early susceptibility measurement by Takeda {\it et al.} 
\cite{cfb_Takeda_74}, the works on CsFeBr$_3$ are devoted to 
the investigation of the magnetic excitations. The excitation spectrum was 
studied both theoretically 
\cite{cfc_cfb_Papanicolaou_89,cfc_cfb_Papanicolaou_90,cfc_cfb_Lindgard_93,cfb_Liu_94} 
and experimentally by means of inelastic neutron scattering in a zero 
field \cite{cfb_Dorner_88,cfb_Dorner_89} and in an external magnetic field 
\cite{cfb_Dorner_90,cfb_Visser_91,cfb_Schmid_92}. 

The lowest frequency excitation mode softens with decreasing temperature 
but stabilizes at 0.11~THz below 2.5~K down to 80~mK \cite{cfb_Schmid_92}.  
This fact indicates that CsFeBr$_3$ remains a SGS system for $T\rightarrow 
0$ in zero field. At 1.6~K in an external magnetic field of 4.1~T applied 
along $c$-axis a well defined Bragg peak appears at (2/3~2/3~1) indicating 
a phase transition to the long range commensurate 120$^\circ$ structure 
\cite{cfb_Dorner_90}. But the correlation lengths do not diverge at that 
field. Instead, they exhibit a flat maximum over about 0.3~T around 4.1~T 
and decrease again at higher field \cite{cfb_Schmid_92}. The nature of 
this phenomena is not yet understood.

\paragraph{RbFeBr$_3$}

This compound can be considered as an intermediate case between the SGS 
antiferromagnet and the Heisenberg antiferromagnet with easy-plane 
 anisotropy; the exchange interaction is strong enough to produce  
three-dimensional order at temperatures below 5.5~K 
\cite{rfb_Eibschutz_73}. At a temperature of 108~K RbFeBr$_3$ undergoes a 
structural phase transition to a distorted phase with space group $P6_3/mmc$ 
\cite{rfb_Harrison_89}, which results in the appearance of two kinds of 
nearest neigbours exchange  in the basal plane ($J'$ and $J'_1$). This 
produces a distortion of a spin triangles with the angle between 
nearest spins not exactly 120$^\circ$.  The spin 
frustration is partially released (see paragraph~\ref{Distort} for details). 
The low temperature crystal phase is found to be ferroelectric \cite{rfb_Mitsui_94}.

The specific 
heat measurements revealed two successive magnetic phase transitions at 
$T_{N1}=5.61$~K and $T_{N2}=2.00$~K \cite{rfb_Adachi_83}. This may be 
caused by the splitting between $J'$ and $J'_1$, but in fact the 
$J'/J'_1$ ratio remains experimentally unknown: the inequivalency of the 
Fe$^{2+}$ sites is not sufficiently large to be distinguished by 
M\"{o}ssbauer spectroscopy \cite{rfb_Lines_75}; the resolution of the 
inelastic neutron scattering experiments \cite{rfb_Harrison_92} was not 
good enough to see the influence of the splitting on the dispersion of 
magnetic excitations.  The energies and intensities of the 
excitations can be described well using the dynamical correlated 
effective-field approximation  neglecting the splitting
 between $J'$ and $J'_1$.

\subsection{The case of ferromagnetic intrachain coupling}
\paragraph{CsFeCl$_3$}

In CsFeCl$_3$ the exchange interaction between Fe$^{2+}$ ions is 
ferromagnetic along the c-axis, while the interchain exchange is weakly 
antiferromagnetic \cite{cfc_Steiner_81}.  The results of inelastic neutron 
scattering \cite{cfc_rfc_Yoshizawa_80} show that at T=5K the lowest 
excitation energy at the magnetic zone center has a peak about 190~GHz 
confirming the absence of long range magnetic order. From the dispersion 
relations the parameters in the effective spin Hamiltonian (\ref{SGS_H}) 
may be obtained, but the results depend strongly upon the theoretical 
model used to analyze the experimental data.  An exciton model, correlated 
effective field  analysis \cite{cfc_rfc_Yoshizawa_80}, 
self-consistent random-phase approximation  
\cite{cfc_cfb_Lindgard_93} and dynamical correlated effective-field 
approximation  \cite{cfc_Suzuki_95} give substantially different 
values for the exchange interaction and the magnetic anisotropy.  The 
parameters of the spin Hamiltonian can be estimated also from measurements 
of the M\"{o}ssbauer effect \cite{cfc_rfc_Montano_74} and of the nuclear 
spin-lattice relaxation time \cite{cfc_Chiba_88} (see Table \ref{SGS_tab} 
for details).

Further inelastic neutron scattering investigation of the dispersion 
curves in the CsFeCl$_3$ \cite{cfc_Schmid_94} has shown that the minimum 
of the dispersion curve does not occur at the $K$-point but is shifted 
slightly away. This effect can be explained by the inclusion of dipolar 
forces \cite{t1_Shiba_82,t2_Shiba_82}.

The transition to a magnetically ordered phase in an external magnetic 
field was detected by  magnetization 
\cite{cfc_rfc_Haseda_81,cfc_Tsuboi_88,cfc_Chiba_88}, specific heat 
\cite{cfc_rfc_Haseda_81}, M\"{o}ssbauer \cite{rfc_cfc_Baines_83} 
and nuclear spin-lattice relaxation \cite{cfc_Chiba_88} measurements.  At 
$H_c=7.5$~T the ordered state was observed at $T<2.6$~K.  Because 
measurements are necessarily made at 
nonzero temperature,  LRO appears over a wide  
region of magnetic field around $H_c$.  For example at $T=1.3$~K 
magnetic susceptibility shows two anomalies at 3.8 and 4.6~T when LRO 
appears, and the same two anomalies at 11.2 and 11.6~T when LRO disappears 
in an increasing field.  The step structure in the $d{\bf M}/d{\bf H}$ 
curve observed at 3.8 and 4.6~T is caused by successive 
phase transitions from the nonmagnetic phase to a thermally frustrated 
incommensurate phase  and then to a commensurate 
three-sublattice antiferromagnetic phase \cite{cfc_Tsuboi_88}.  An 
incommensurate magnetic phase (double-modulated and single-modulated) 
was found between the nonmagnetic phase at low field and the commensurate 
phase at higher 
field by means of elastic neutron scattering 
\cite{cfc_Knop_83}. At T=0.7~K the magnetic phase transitions take place 
at $H_1=3.85$~T, $H_2=3.92$~T and $H_3=4.5$~T.  Possible explanation of 
the nature of the incommensurate phases has been given in a framework of 
the correlation theory \cite{cfc_Lindgard_86,cfc_cfb_Lindgard_93}.

An additional magnetic phase transition at 33~T was observed in a 
magnetization measurements \cite{cfc_Tsuboi_88}.  The high field 
magnetization cannot be explained within the framework of the fictitious $S=1$ 
spin states. Since the magnitude of themagnetization at 33 T is large, the anomalous 
increase in ${\bf M}$ is attributed to the upper excited ${\cal J}=2$ spin 
state \cite{cfc_Tsuboi_88}.

All five possible transitions for $H\parallel c$ and $H \perp c$ between 
the ground state and the excited doublet (see Fig.\ref{sgs_h_split}) have 
been observed in submillimetre wave ESR-measurements \cite{cfc_Ohta_92_1}, 
while at higher frequencies only two absorption lines were observed using 
far infrared Fourier spectroscopy \cite{cfc_Ohta_92_2}.  One of the 
absorption seems to come from the excitation between ground state and 
second excited doublet. The mechanism of the second absorption is still 
unclear.

\paragraph{RbFeCl$_3$}

Unlike CsFeCl$_3$ the isomorphous compound RbFeCl$_3$ reveals  
three-dimensional long range magnetic order below $T_{N1}=2.55$K 
\cite{???_Davidson_71} The signs of the exchange interactions are the 
same as in CsFeCl$_3$ -- ferromagnetic along the c-axis and 
antiferromagnetic in the plane. An inelastic neutron 
scattering study \cite{cfc_rfc_Yoshizawa_80,rfc_Yoshizawa_81} showed clear 
softening of the magnetic excitations in a small region around the zone 
center when $T \rightarrow T_{N1}$.  The influence of the  
softening on the nuclear spin-relaxation time $T_1$ of $^{87}$Rb was 
observed in an NMR experiment \cite{rfc_Goto_86}.  According to an 
elastic neutron scattering study \cite{rfc_Wada_82} at zero magnetic field, 
RbFeCl$_3$ undergoes three transitions at $T_{N1}=2.5$~K, $T_{N2}=2.35$~K 
and $T_{N3}=1.95$~K.  Two different incommensurate structures have been 
found at $T_{N3}<T<T_{N2}$ and $T_{N2}<T<T_{N1}$, while at $T<T_{N3}$ the 
 120$^\circ$ in-plane triangular structure is observed 
 \cite{???_Davidson_71,cfc_rfc_Yoshizawa_80,rfc_Wada_82}.  Very similar 
values for the phase transition temperatures have been found from specific 
heat and susceptibility measurements \cite{cfc_rfc_Haseda_81}. The 
existence of thermal hysteresis at $T_{N3}$ indicates that the 
incommensurate-commensurate transition is  first order. From this study and 
susceptibility measurements \cite{rfc_Wada_83} a phase diagram in $(H,T)$ coordinates may 
be derived as shown in Fig.~\ref{RbFeCl3_PhD}. There are two main features a) 
the 120$^\circ$-structure and paramagnetic phase are always separated by 
one or two incommensurate phases; b) application of a magnetic field 
parallel to the $c$-axis increases the phase-transition temperature until
 a maximum is reached at  7.5~T; subsequently there is a rapid decrease 
and the transition temperature goes to zero at 13~T. Shiba and 
Suzuki~\cite{t2_Shiba_82} have proposed a theory, which explain the 
experimental phase diagram reasonably well from the viewpoint 
of a conical-point instability due to the dipole-dipole interaction.  
They show that even a small dipole-dipole 
interaction can transform the 120$^\circ$ structure to an incommensurate 
structure at intermediate temperatures, although the low 
temperature phase still should have the 120$^\circ$ structure. Very recently 
the phase diagram of RbFeCl$_3$ for $H \perp c$ was reinvestigated by 
neutron scattering  \cite{rfb_Dorner_95}. Good agreement 
between the results of two experiments was found for a field less than 
1.0~T. Above 1.0~T the neutron scattering study gives a slightly higher value 
of the transition field from the commensurate phase to the incommensurate 
phase. 

The high field magnetization of RbFeCl$_3$ exhibits an anomaly around 31~T 
\cite{rfc_Amaya_88}, similar to those in CsFeCl$_3$ \cite{cfc_Tsuboi_88}.
The situation regarding the spin Hamiltonian parameters derived from 
experiment for  RbFeCl$_3$ is very similar to that in CsFeCl$_3$. Different 
authors analyzed their experimental data on the basis of different 
approximations and there has been controversy about the value of the 
exchange interactions and the anisotropy. The results of a susceptibility 
measurements were analyzed using the molecular field approximation  
\cite{multi_Achiwa_69} or pair approximation \cite{??_Montano-73}, where a 
Fe$^{2+}$ chain was represented by an assembly of isolated pairs of 
nearest neighbour spins. The M{\" o}ssbauer and susceptibility data 
\cite{rfc_Eibschutz_75} were analyzed using the correlated effective-field 
approximation, developed by Lines \cite{t_Lines_75}. The results of 
inelastic neutron scattering \cite{cfc_rfc_Yoshizawa_80} were analyzed 
using this theory, the three sublattice spin-wave approximation and the exciton 
model. The parameters so determined depend strongly on which approximation 
is used;  
they are not universal;  one set of parameters cannot describe all 
available experimental data. Suzuki \cite{rfc_Suzuki_81} has made an 
attempt to see to what extent one can understand the various observed 
magnetic properties on the basis of a single set of parameters. He  
used thedynamical correlated effective-field approximation 
\cite{??_Suzuki_78} and has found values of $D, J_\perp, J_\parallel, g_\perp$ and 
$g_\parallel$ which can reproduce experimental data reasonably well. 
Originally  he has considered for the sake of simplicity only a single 
chain of Fe$^{2+}$ ions, so that the approach has applicable only in 
the paramagnetic phase. Later he has 
included an interchain coupling in the model and derived a 
consistent set of exchange parameters which explain the behaviour of 
RbFeCl$_3$ above and below $T_N$ \cite{rfc_Suzuki_83,t_Suzuki_83}.

\section{Triangular antiferromagnet stacked ferromagnetically}
\label{Ferro}
Except for  CsFeCl$_3$ and RbFeCl$_3$ as described in the previous chapter, 
there are just two triangular antiferromagnets with ferromagnetic 
interactions along the chains, CsCuCl$_3$ and CsNiF$_3$. Each gives rise 
to its own different physics, and will be dealt with in separate subsections 
of this chapter.

\subsection{CsCuCl$_3$}

CsCuCl$_3$ has been one of the most extensively studied of the triangular
antiferromagnets.  The spins in the copper chains are coupled
ferromagnetically, but the planar interactions are antiferromagnetic so that
frustration effects are of similar importance to the common case of
antiferromagnetic chains.

What makes CsCuCl$_3$ unique is that below 423~K the triangular crystal
structure is distorted through the Jahn-Teller effect to give a crystal
structure with space group $P6_{1}22$ \cite{ccuc_Schlueter_66,ccuc_Kroese_74}.
The distortion from the stacked triangular lattice involves small in-plane
displacements of copper atoms so as to form a helix with axis along $c$ with 
one turn of the helix every six layers.

Below $T_N$=10.66(1)~K \cite{ccuc_Mekata_95} the zero-field magnetic structure
shows $ab$ planes with the $120^{\circ}$ triangular antiferromagnetic structure
and with moments in the plane.  Magnetic neutron Bragg peaks are observed at
($\frac{1}{3} \; \frac{1}{3} \; 6n{\pm}{\delta}$) with $\delta$=0.085
\cite{ccuc_Adachi_80}. This corresponds to the triangular spin arrays that are
 rotated about c by $5.1^{\circ}$ between successive planes. The period of 
 rotation is 11.8c or 214 \AA.

It is believed \cite{ccuc_Adachi_80} that this rotation arises from the 
Dzyaloshinsky-Moriya (DM) interaction that gives rise to a term in the Hamiltonian 
given by
\large
\begin{equation}
{\cal{H}}_{DM} = \sum_{i,j} {\bf D}_{ij} {\:} \cdot {\:} ({\bf S}_i \times {\bf S}_j).
\label{DM_term}
\end{equation}
\normalsize
In CsCuCl$_3$ ${\bf D}_{ij}$ is a vector along $c$ which is non zero only when
atoms $i$ and $j$ are nearest neighbours along the helical chains.  In the
 absence of the Jahn-Teller distortion, symmetry requires the DM term to vanish;
 this is why it has not been included in the Hamiltonian for other materials
 discussed in this work.  The Jahn-Teller distortion gives a second, smaller, 
 effect on the Hamiltonian in that it causes the vector ${\bf{D}}_{ij}$ to deviate 
 slightly from from the c axis with a period of six lattice 
 spacings~\cite{ccuc_Adachi_80,ccc_Tanaka_85}.  This effect is small and often 
 neglected in the literature.   Neutron scattering measurements of the spin-wave 
 dispersion relations by
 Mekata {\it et al.}~\cite{ccuc_Mekata_95} confirm this Hamiltonian and give the
intra-chain exchange $J=-580$~GHz, $|D|=121$~GHz and the in-plane exchange
interaction $J'=97$~GHz (no quantum corrections were included in deriving these 
parameters). It is 
apparent that CsCuCl$_3$ is less one
dimensional than is usual for ABX$_3$ materials, since $|J/J'|$ is about 6
and in other materials it is one to two orders of magnitude larger.

The ferromagnetic interactions along the chain give a minimum energy when all
the moments are aligned parallel, but the DM interaction is minimised when
neighbouring moments are aligned perpendicularly.  The sum of these two terms
in the Hamiltonian gives a minimum energy classically for a helical magnet
with the tangent of the turn angle between neighbouring moments equal to 
$|D/2J|$~\cite{ccuc_Adachi_80}. The observed magnetic structure corresponds 
to these helices along $c$ stacked on a $120^\circ$ triangular lattice. The
 parameter $\delta$ is observed to be independent of temperature \cite{ccuc_Adachi_80} 
 and field ($H<H_c$) \cite{ccuc_Stuesser_95}.  Spin-wave dispersion relations have 
 been calculated by Rastelli and Tassi
\cite{ccuc_Rastelli_94} and by Stefanovskii and Sukstanskii
\cite{ccuc_Stefanovskii_93} and shown to agree with antiferromagnetic resonance
data. 

The maximum ordered moment for $S=\frac{1}{2}$ and g=2.10 \cite{ccuc_Ohta_93}
is 1.05~$\mu_B$.  Adachi {\it et al.} \cite{ccuc_Adachi_80} reported
the ordered moment as $(0.61 \pm 0.01)$~$\mu_B$, extrapolated to 0~K, but 
recent work gives higher values, 0.85~$\mu_B$ \cite{ccuc_Mekata_95} for zero field 
and low
temperature, and $(0.90 \pm 0.01)$~$\mu_B$ at $T=10.35$~K and $H=5.9$~T
\cite{ccuc_Stuesser_95}.  The two latter values are higher than is typical for
frustrated systems, perhaps because the ferromagnetic chain coupling leads to
smaller moment reductions than does antiferromagnetic coupling.

An applied field along $c$ gives rise to two first-order phase transitions at
fields $H_{c1}$ and $H_{c2}$. Nojiri { \it et al.} \cite{ccuc_Nojiri_88} 
gives $H_{c1}=12.5$~T and $H_{c2}=31$~T at $T=1.1$~K, while Chiba 
{\it et al.}~\cite{ccuc_Chiba_95} gives $H_{c1}=11.19$~T at $T=4.2$~K. For fields 
less than $H_{c1}$ the magnetic structure is that found in zero field together with
a canting of each spin towards the direction of the applied magnetic field. At
$H_{c1}$ the triangular layers break down into a colinear structure on the
triangular lattice based on that shown in figure \ref{ising_spin_arr}, where two 
spins are aligned
in one direction and the third is in the opposite direction.  There is a
canting of all three spins towards the field direction \cite{ccuc_Nikuni_93}. 
The helical stacking of the planes remains in this structure. Neutron
scattering work \cite{ccuc_Stuesser_95} confirms this description of the
magnetic structure above $H_{c1}$.  On passing from the low-field to the
medium-field structure the ($\frac{n}{3} \; \frac{n}{3} \; 6l\pm \delta$) lines
lose intensity and new lines appear at ($\frac{n}{3} \; \frac{n}{3} \; 6l$).  
Above $H_{c2}$
the magnetic structure is believed to be almost ferromagnetic, with the helical
stacking destroyed.  The phase transition at H$_{c1}$ is not predicted for a
classical system with the appropriate Hamiltonian;  Nikuni and Shiba
\cite{ccuc_Nikuni_93} show however that when quantum fluctuations are taken
into account the phase transition is to be expected.

The field H$_{c1}$ decreases slowly as the temperature 
increases~\cite{ccuc_Fedoseeva_85,ccuc_Gekht_89,ccuc_Stuesser_95,ccc_Weber_95}, 
and this line of first-order phase transition in the H-T phase diagram meets 
the paramagnetic phase transition at a  bicritical point. Extrapolation of the 
neutron scattering measurements of Stuesser {\it et al.} \cite{ccuc_Stuesser_95} 
indicate that the bicritical point occurs at $H_{B}=5.5$~T and $T_{B} = 10.7$~K, 
while the specific heat measurements of Weber {\it et al.} \cite{ccc_Weber_95} 
give the same value for H$_{B}$ and T$_B$ = 10.59~K.

The behaviour of CsCuCl$_3$ in a field perpendicular to the $c$ axis has been
 treated theoretically by Jacobs {\it et al.} \cite{ccuc_Jacobs_93,ccuc_Ohyama_95}. 
 The field splits the degeneracy of the orientation of the spin triangle in the 
 $ab$ plane and causes the triangles to no longer have angles of $120^{\circ}$.  
 This results in $\delta$ varying with the applied field $H$.  An anomaly is 
 observed \cite{ccuc_Nojiri_88} in the low-temperature magnetization at 12 T 
 which is believed to involve a transition to a commensurate state.  Since 
 CsCuCl$_3$ is a frustrated system with S=$\frac{1}{2}$, quantum fluctuations 
 would be expected to be important.  Jacobs {\it et al.} \cite{ccuc_Jacobs_93} 
 confirm this by evaluating the first term in a $1/S$ expansion and showing that 
 quantum (and thermal) fluctuations lift a nontrivial degeneracy and stabilize 
 the commensurate state.

We finish this subsection by describing work on the critical phase transition.  
The magnetism in CsCuCl$_3$ shows two chiral degeneracies, one arising from the 
triangular structure in each plane and one from the helix along the $c$ axis.  
Weber {\it et al.} \cite{ccc_Weber_95} argue that the structural helix will 
not affect the critical properties and that the phase transition should be 
that of the chiral XY model.  This model predicts $\beta = 0.25 \pm 0.01$ and 
$\alpha = 0.34 \pm 0.06$ (table V), while if the non-universality model holds 
tricritical exponents would be observed with $\beta = 0.25$ and $\alpha = 0.5$ 
(table III).
The early neutron-scattering measurements of Adachi {\it et al.}
\cite{ccuc_Adachi_80} gave $\beta =0.358 \pm 0.015$, but this value has not been
confirmed by recent work that gives much lower values; Mekata
{\it et al.} \cite{ccuc_Mekata_95} give $0.25 \pm 0.01$ and Stuesser 
{\it et al.}~\cite{ccuc_Stuesser_95} give $0.23 \pm 0.02$. Recent specific heat 
measurements of Weber {\it et al.} \cite{ccc_Weber_95} in zero field can be 
described well by a critical exponent $\alpha = 0.35 \pm 0.05$ except very close  
to T$_N$ ($t<10^{-3}$) where the transition seems to go over to being weakly 
first order. The exponent is compatible with the chiral XY model and not with 
the tricritical model.  The small canting of the vector ${\bf{D}}_{ij}$ from the 
$c$ axis will lower the symmetry from  $Z_{2}{\times}S_{1}$ to $Z_{2}$. Thus the 
critical properties should eventually exhibit a crossover effect from 
$Z_{2}{\times}S_{1}$ to $Z_{2}$ behavior, though this effect has not been observed 
to date. It is not clear whether the observed weakly-first-order effects are a 
fundamental  property of  the frustrated $XY$  model or if they are a consequence 
of the expected crossover.

As the field increases towards H$_B$ the amplitude of the specific heat divergence 
becomes smaller and the first order part of the transition becomes even weaker. 
At a field above H$_B$, $H = 7 T$, the specific heat measurements show a critical 
phase transition with $\alpha = 0.23 \pm 0.08$, contrary to the expected value 
for regular XY behaviour where $\alpha = -0.01$.  The origin of this discrepancy 
is not clear.

\subsection {CsNiF$_3$}

This is the only fluoride ABX$_3$ compound which crystallises with the stacked
triangular lattice \cite{cnf_Babel_69}.  The magnetic structure corresponds to
an easy-plane antiferromagnet with no distortions at low temperatures.  As
usual the interactions are much stronger along $c$ than within the $ab$ plane,
giving rise to quasi-one-dimensional properties above $T_N$.  The interactions
along the chain and the soliton properties have been much studied in
the quasi-one-dimensional temperature region (see Kakurai {\it et al.} 
\cite{cnf_Kakurai_84} and references therein).  In this article
only the three-dimensional ordered properties will be discussed.  These are
unique for an easy-axis material since the ordering consists of ferromagnetic
$ac$ planes stacked antiferromagnetically with moments aligned along $a$
\cite{cnf_Steiner_72}, as shown in the figure~\ref{CsNiF3}.

This ordering implies that the low-temperature ordered state breaks the
hexagonal symmetry.  In practice domains are formed favouring one of the three
equivalent a directions in the basal plane.  In terms of the hexagonal unit
cell, Bragg peaks are observed in the low-temperature neutron-scattering
pattern with indices (h/2,k,l) where h, k and l are all integers.

The magnetic order cannot arise from a Hamiltonian containing just nearest-neighbour
interactions and easy-plane anisotropy, as has been shown in earlier chapters.
Scherer and Barjhoux \cite{cnf_Scherer_77} and Suzuki \cite{cnf_Suzuki_83} show
that it will occur if the interactions, other than those along the chains, are
predominantly dipolar in character.  The idea of this predominance is supported
by the low value of the  N\'{e}el temperature, 2.7~K, which shows weak interchain
interactions. The value of the ordered moment, 2.26~${\mu}_B$
\cite{cnf_Steiner_72}, for S=1 and g=2.28, is higher than is found in
frustrated triangular structures. A theoretical treatment which includes
magnetoelastic effects has been given by Caill\'{e} and Plumer.  A mean-field
treatment of the phase diagram is given by Trudeau and Plumer.  

The various experimentally-determined values of $J$, $J'$, $D$ and $T_N$ are listed
in the table~\ref{cnf}.  It is clear that the in-plane exchange, $J'$, is small
compared with the exchange along the chains, $J$, but the magnitude is similar to
that found in other ABX$_3$ easy-plane materials (see table
\ref{Heis_parameters}). The anisotropy, $D$, is however larger than in other
easy-axis materials. The dipolar forces are long range in nature and their
relative influence, which comes from sums over a large number of moments, is
greater for ferromagnetic than for antiferromagnetic chain interactions.
Theory \cite{cnf_Scherer_77,cnf_Suzuki_83,cnf_Plumer_88} predicts the correct
structure for the parameters given in the table together with dipolar
interactions.

Because the chains are ferromagnetic, it would be expected that applied
magnetic fields will have a relatively greater influence than in other ABX$_3$
materials.  This is indeed the case for fields applied in the basal plane,
though the strong easy-plane anisotropy makes the effect of fields applied
along c less.

The neutron scattering measurements of Steiner and Dachs \cite{cnf_Steiner_74}
show that small fields, $H$, applied in the basal plane to the ordered material
influence the relative sizes of the three domains.  A field of around 0.05~T is
sufficient to produce a single-domain sample with moments approximately
perpendicular to $H$.  Larger fields destabilize the antiferromagnetic state and
there is a phase transition at critical field, $H_c$, to a paramagnetic state
with imposed alignment of the moments along H.  $H_c$ is around 0.2 to 0.3~T at
2.0~K \cite{cnf_Steiner_74,cnf_Lussier_93}

If the field is applied along $c$, larger values of $H$ are needed to destroy the
antiferromagnetism because the field is opposed by the easy-plane anisotropy. 
A similar phase transition occurs in this case, but the critical field, $H_c$
is 8 to 25 times larger than when $H$ is applied in the basal plane
\cite{cnf_Lussier_93,cnf_Plumer_88}.

The critical properties of CsNiF$_3$ are not simple.  The Hamiltonian has XY
symmetry in zero field or with applied field $H<H_c$ along $c$. For an applied
field in the plane sufficient to produce a single-domain sample the symmetry
becomes Ising like \cite{t_Lovelock_77} with $z$ in a direction perpendicular
 to both $c$ and $H$. There
are two complicating factors however. First the long-range nature of the
dipolar interactions can lead to mean-field exponents and to crossover
behaviour.   Second the presence of three equivalent domains in the structure
changes the critical properties and makes the phase transition first order
\cite{t_Loisson_93}.

Neutron scattering measurements in zero field \cite{cnf_Steiner_80} show a
critical phase transition with a crossover at ${\epsilon}=|T-T_{c}|/T_{c}$
values near $1.3\times10^{-2}$ from exponents, $\gamma = 1.1 \pm 0.1$ and
$\nu = 0.54 \pm 0.07$, at large $\epsilon$, to different values, $\gamma =
1.45 \pm 0.10$, $\nu  = 0.68 \pm 0.07$ and $\beta = 0.34 \pm 0.04$, at small
$\epsilon$.  Although the errors on the exponents are not small, the first set
is consistent with mean-field exponents and the second set with XY behaviour.
There is no experimental evidence for a first-order phase transition. Strain
effects have been observed in CsNiF$_3$ crystals \cite{cnf_Lussier_93} which
might suppress domain fluctuations, but this would not be expected to
completely change the nature of the transition.

There is one other set of measurements of critical properties.  Lussier and
Poirier \cite{cnf_Lussier_93} have measured the phase boundary $H_c$ as a
function of temperature and of the direction of $H$.  For fields perpendicular
 to $c$, $H_c$ is large enough for the sample to be single domain and Ising
 symmetry is to be expected.  Plumer and Caill\'{e} \cite{cnf_Plumer_88} show 
 that $H_c$ varies as the order parameter, so that the measurements should 
result in an
Ising critical exponent $\beta$=0.326. The measured exponent is 0.31$\pm$0.01 in
not-too-bad agreement with theoretical expectations.
The situation with $H$ along $c$ is puzzling however. Theory predicts an XY
exponent, $\beta$=0.345, for H$_c$ while the experiment gives a higher value,
$\beta$=0.37$\pm$0.01.  

\section{Diluted and mixed triangular magnets}
\label{Impurity}

It is apparent from previous chapters (or at least authors hope it is apparent)
that there is a plenty of triangular magnets, physical properties of which
depend on type and relative strength of exchange and anisotropic interactions.
Those properties are not totally established yet.  Even less understanding of
properties of diluted and mixed triangular magnets is currently achieved --
they found to be sophisticated as much again.  Nevertheless, some interesting
results and ideas have been found in the process of their investigation. 
It is useful to draw the analogy here between the triangular antiferromagnets
and another example of frustrated magnetic system, XY square-lattice
antiferromagnet dominated by second-neighbor antiferromagnetic exchange.
In the latter case dilution acts against thermal and quantum fluctuations,
producing an effect known as "ordering due to disorder"~\cite{theor_Henley_89}.
Present chapter summarises briefly characteristics of diluted and mixed 
magnets on triangular lattice.

The obvious method to "disturb" magnetic system is to introduce small amount 
of nonmagnetic impurities.  The presence of an impurity results either 
in mechanical distortion of the structure of the original crystal or in disruption 
of some part of the interaction between the magnetic atoms, which 
is reflected in the collective behaviour of the spin system.  Random field
effect due to non-magnetic impurities on spin correlation  was studied by
elastic neutron scattering and magnetic susceptibility measurements in 
 Ising antiferromagnet, CsCoCl$_3$, doped by Mg or Zn \cite{ccocl_d_Mekata_87} 
 and by measurements of diffuse scattering in 
CsCoCl$_3$ doped by Mg \cite{ccoc_d_Mekata_90}.  The reduction of the upper
magnetic transition temperature, $T_{N1}$, has been found in samples with
impurities: $T_{N1}=21.0$~K, $T_{N1}=20.3$~K and $T_{N1}=19.8$~K for the 
pure crystal, crystal with 0.58\% of Mg and crystal with 1.7\% of Mg 
respectively \cite{ccoc_d_Mekata_90}, while temperature of the lower magnetic
 transition, $T_{N2}$, could not be found down to 1.6~K \cite{ccocl_d_Mekata_87}. 
 The Mg concentration was determined by a chemical
 analysis.

On the other hand, detailed ESR and magnetization measurements of 
 Heisenberg triangular antiferromagnet, RbNiCl$_3$, doped with 1\% of Mg
 showed that the Neel temperature remained unchanged from pure crystal, in 
which $T_N \approx 11$~K  \cite{rnc_Zhitomirsky_95}. However, the influence 
of the impurity at low temperature is still well pronounced:  1) the spin-flop
region became much broader, 2) the gap $\omega (H=0)$ of one of the resonance
branches has increased from 55~GHz to 61~GHz, 3) the gap of another resonance
branch has found to be 20~GHz, while in the pure crystal it has not been
observed and the estimated value for pure crystal is 0.2~GHz. Such a dramatic
changing of the resonance spectrum was successfully described by introducing 
a two-ion anisotropy of the form $D(S_i^z)^2(S_j^z)^2$ into the spin
Hamiltonian~(\ref{H_H}). It has been postulated that an impurity which does 
not occupy a site in the crystal lattice strongly distorts the electrical
interactions within the crystal, altering the character of the anisotropic
interactions.  In contrast to the case of doped CsCoCl$_3$, where Mg
concentration was different in different crystals, a $\gamma$~activation
analysis reviled approximately the same concentration (about 1\%) of Mg in 
all investigated crystals of RbNiCl$_3$.

Very unusual result has been reported recently by Yamazaki
{\it et al.}~\cite{csvcl_del_Yamazaki_96} for a CsV$_{1-x}$Mg$_x$Cl$_3$,
($x$=0.000 - 0.357).  Temperature-dependence of the magnetic susceptibility
for a sample with $x$=0.026 suggests that the ordering temperature is about
35~K, while in a pure sample $T_N1=13.8$~K~\cite{avx_Hauser_85}. Such
a dramatic increase of the ordering temperature obviously has to be
confirmed by some other techniques.

The influence of diamagnetic dilution on the magnetic ordering process of the 
induced-moment antiferromagnet RbFeCl$_3$ was studied in 
ref.~\cite{rfc_d_Harrison_90}.  Single crystals of the solid solution 
RbFe$_{1-x}$Mg$_x$Cl$_3$ ($x$=0.02, 0.03 and 0.05) were investigated by means 
of elastic neutron scattering.  The $x$=0.02 and 0.03 samples showed 
transitions from paramagnetic to the IC$_1$ phase at the same temperature,
$T_{N1}=2.55$~K (see fig.~\ref{RbFeCl3_PhD}), then transitions to the IC$_2$ 
and C phases at temperatures that decreased sharply with $x$. The $x$=0.05
sample also shows a transition to the IC$_1$ phase at $T_{N1}=2.55$~K, but 
no further transitions down to the lowest temperature of 1.38~K.  
An additional elastic diffuse magnetic scattering component centred at the
vector $(\frac{1}{3} \frac{1}{3} 0)$ has been found to persist to temperatures
well above $T_N$ in all samples. 

Another method of investigation of the magnetic system consists of introducing
small amount of magnetic impurities. A substitution of magnetic ions can
modify the amplitude, or even sign, of the effective single-ion anisotropy.
For example, in pure RbNiCl$_3$ the splitting between $T_{N1}$ and $T_{N2}$
is very small, 0.14~T \cite{rnc_Oohara_91_1}, while addition of only 5\% of Co
results in as much as 9~K between $T_{N1}$ and $T_{N2}$ \cite{rnc_d_Oohara_94}.
Magnetic phase diagrams of Heisenberg triangular antiferromagnet,
CsNi$_{0.98}M_{0.02}$Cl$_3$ ($M$=Co, Fe, Mg) have been determined by heat
capacity \cite{cnc_d_Takeuchi_93} and ultrasonic velocity \cite{cnc_d_Trudeau_95} 
measurements. As expected, when comparing with pure CsNiCl$_3$,  the Co-doped 
crystal shows enhanced Ising effective single-ion
anisotropy --  the spin-flop field $H_{SF}$ and the splitting between $T_{N1}$
and $T_{N2}$ are sufficiently increased; the Fe-doped crystal behaves as a 
Heisenberg antiferromagnet with $XY$ type of anisotropy  -- the phase diagram 
is very similar to those in CsMnBr$_3$; in the Mg-doped crystal nonmagnetic
impurity causes decrease of effective single-ion anisotropy and values of
$H_{SF}$ and $T_{N}$ consequently. Magnetic phase diagrams of 
CsNi$_{0.98}M_{0.02}$Cl$_3$ for $H \parallel c$ is shown on fig.~\ref{deluted}.

Apart from addition of small amount of impurities into antiferromagnet on
triangular lattice, some investigations had been done on mixed systems, 
where the impurity concentration is not small. In ref.~\cite{rnc_knc_Tanaka_93}
the magnetic phase diagram of Rb$_{1-x}$K$_x$NiCl$_3$ has been studied by
susceptibility and torque measurements. Pure RbNiCl$_3$ is a Heisenberg
antiferromagnet with easy-axis type anisotropy, while KNiCl$_3$ demonstrates
large easy-plane type anisotropy, see Chapter~\ref{Heis}. The transition 
between two different types of anisotropy has been found at $x_c=0.38$.  
The interpretation of the observed phase diagram at large $x$ is quite
complicated due to crystal structure distortions of pure KNiCl$_3$.

A.Harrison and coauthors \cite{rfc_cfc_Harrison_86,rfc_rfb_Harrison_89,simul_Harrison_89} 
had investigated magnetic ordering effects in the mixed singlet-groud-state 
magnets AFeX$_3$, where A is a mixture of Cs and Rb or X is the mixture of 
Cl and Br. As it was shown in the Chapter~\ref{SGS} CsFeCl$_3$ has "truly"
nonmagnetic ground state, while in RbFeCl$_3$ sufficiently strong exchange
interaction causes magnetic ordering at $T<T_N=2.5$~K. Therefore, the
replacement of Rb in RbFeCl$_3$ by Cs should decrease Neel temperature or even
suppress magnetic ordering. Indeed, it was found experimentaly \cite{rfc_cfc_Harrison_86}, 
that as low as 5\% of Cs is sufficient to destroy
long-range magnetic ordering. In RbFeCl$_{3-x}$Br$_x$ there is 
an obvious competition between ferro and antiferromagnetic sign of the intrachain 
interactions: in RbFeCl$_3$ the exchange along $c$-axis is
ferromagnetic, while in RbFeBr$_3$ it is antiferromagnetic (see Chapter~\ref{SGS}).  
Again, low concentrations of either type of dopant, $0.3<x<2.7$, destroys the
magnetic long-range order \cite{rfc_rfb_Harrison_89}. Note, that at 
intermediate compositions a singlet ground-state phase has been observed, 
rather than expected a spin-glass phase.   

\section{Conclusions}
\label{Concl}

As it was mentioned in the Introduction, there are several ways that 
triangles of antiferromagnetic interactions can be built into a crystal 
lattice. The distinguished feature of the majority of antiferromagnets 
on a stacked triangular lattice, described in this article, is the 
appearance at sufficiently low temperature of the three-dimensional 
long-range magnetic ordering, despite the frustration of the exchange 
interaction. The transition temperature of this sort of antiferromagnets 
is typically an order of magnitude lower than the Curie-Weiss temperature, 
but such a big reduction of the transition temperature usually arises from 
a combination of two effects,  frustration  and the low-dimensionality of 
 actual stacked-triangular materials  (they have quasi 1D or 2D character).

The influence of the frustration on magnetic properties of other triangular 
antiferromagnets, which have more complicated structures, is apparently more 
substantial. For example, in Kagome-lattice antiferromagnet 
SrCr$_{9p}$Ga$_{12-9p}$O$_{19}$, with $p\approx0.9$~\cite{kagome}, 
the low-temperature ground state is a spin glass. In gadolinium gallium 
garnet, Gd$_3$Ga$_5$O$_{12}$, where the magnetic Gd ions are on two 
interpenetrating corner-sharing triangular sublattices, long range magnetic 
order has been found only in the applied magnetic field around 
1~T~\cite{GGG_field}, while in a lower field magnetization is different 
for a field cooling and zero field cooling~\cite{GGG}, which is typical 
for a spin glass. Many pyrochlores with the chemical formula
A$_2$B$_2$O$_7$ also undergo a phase transition to a spin glass 
state~\cite{Gaulin_94}.

Because triangular antiferromagnets on a stacked triangular lattice still 
can  be described at low temperature in terms of Neel-type ordering, rather than spin-glass 
or a short-range order, they form a good basis for the study of the effects 
of frustration on magnetic systems. Frustration leads to new physics with 
novel phase diagrams and critical properties. Zero-point fluctuations are 
found in Ising systems; these become large in $XY$ and Heisenberg systems.

Theory and experiment seem to be in good accord in describing the ground
state and the excitations of the triangular antiferromagnets except in the
case of S=1 quasi-one-dimensional nickel materials where spin-wave theory
seems to be inadequate. Correction terms for
ground-state fluctuations are larger than can be dealt with confidently and
vestiges of the one-dimensional Haldane effect are believed to be present.

Theoretical predictions of the nature of the phase diagram as a function of
$H$ and $T$ are in accord with experiment, but the critical properties at
phase transitions are often not described satisfactorily.  For materials
with the Heisenberg Hamiltonian, the theoretical consensus favours Kawamura's
SO(3) chiral universality class, but experiments show significant
discrepancies. For the $XY$ Hamiltonian the theoretical situation is
controversial with three contending scenarios, chiral $XY$ properties,
tricritical properties and a weak first-order phase transition. Experiment,
which is largely confined to one material, CsMnBr$_3$, shows a critical
phase transition with exponents that can be taken to be in agreement with
either the chiral $XY$ model or tricritical exponents; the two theories
give quite similar predictions for the exponents.  There is a need for 
measurements of critical exponents of weak ($D<3J$') easy-plane materials 
and for strong easy-plane materials in a field.

For easy-axis materials the experimental values and the theoretical 
predictions for the critical indices $\beta$, $\nu$ and $\gamma$ at both 
of the two zero-field phase transitions are irreconcilable. The experimental 
values are not in accord with the scaling laws and a confirmation of the 
single report of values of $\nu$ and $\gamma$ would be desirable.

A number of cobalt quasi-one-dimensional triangular antiferromagnets have
Hamiltonians that are close to the Ising Hamiltonian.  The zero-field
properties of these materials is not simple with three phase transitions.
Only the upper one of these shows a specific-heat anomaly. Most theoretical
works have predicted critical exponents at the upper critical temperature
that follow the three-dimensional unfrustrated $XY$ model, and most
experiments agree with these predictions. One recent Monte-Carlo study
suggested slightly different exponents, but unfortunately the experiments
are not sufficiently accurate to distinguish. One group has recently
reported experimental determinations of $\beta$ that are significantly
lower than those given in other reports, and that are not in accord with
any theoretical predictions.

It has long been a puzzle why one triangular antiferromagnet, RbMnBr$_3$ 
shows an incommensurate magnetic structure. Recently we and our coworkers 
have shown this to arise from small structural distortions of the lattice 
along planes perpendicular to the basal plane. These distortions, which map 
on to the row model of Zhang, Saslow and Gabay, allow relatively small 
changes in exchange parameters, arising from the distortions, to destroy 
the simple triangular magnetic structure. Similar distortion effects are 
also found in KNiCl$_3$.

\acknowledgments

We are grateful to Drs. R. Feyerherm, B.D. Gaulin, J. Gardner, M.L. Plumer,
D. Visser and M. Zhitomirsky for useful discussions
about some aspects of this manuscript. We thank the Natural Sciences
and Engineering Research Council of Canada for financial support. One of
the authors (MFC) wishes to thank the Chalk River Laboratories for
hospitality during a sabbatical visit while the this article was written.

\newpage
\begin{table}[p]
\begin{center}
\begin{tabular}{clccccrc}
{\bf A:B}& Co & Cu & Fe & Mn & Ni & V &{\bf X}\\ \hline
Rb &Ising\cite{ccoc_ccob_rcol_Hori_90} &non Hex\cite{???_Crama_81}& 
F or SGS\cite{cfc_rfc_Yoshizawa_80}&non Tr\cite{rmc_Melamud_71}&
HEA\cite{rnc_Yelon_72}& HEP\cite{avx_Hauser_85}& Cl \\
Cs &Ising\cite{ccoc_Mekata_78} & F\cite{ccuc_Adachi_80} & SGS
\cite{cfc_rfc_Yoshizawa_80}&non Tr\cite{cmc_Melamud_71}&
HEA\cite{rnc_cnc_Minkiewicz_70}&HEP\cite{avx_Hauser_85}& \\
Rb &Ising\cite{rcb_rnc_Minkiewicz_71} & - &HEP or SGS\cite{rfb_Harrison_92}& 
HEP \cite{rmb_Glinka_73} &HEA\cite{abx_Witteveen_74}&HEP\cite{avx_Hauser_85}&Br \\
Cs &Ising\cite{ccob_Yelon_75}& - & SGS\cite{cfb_Dorner_88}& 
HEP\cite{cmb_Eibshutz_72} &HEA\cite{cnb_Brener_77}&HEP\cite{avx_Hauser_85}& \\
Rb & - & - & - & - & - &HEP\cite{abi_Zandbergen_81}& I \\
Cs & - & - &HEP or SGS\cite{cfi_Zandbergen_81}& HEA\cite{cmi_Zandbergen_80} &
$^\star$&HEP\cite{avx_Hauser_85}& \\
\end{tabular}
\end{center}

\vspace*{1 mm}
\noindent
KNiCl$_3$ -- HEP\cite{knc_Tanaka_89,knc_Petrenko_95},
CsNiF$_3$ -- F\cite{cnf_Steiner_71}.

\vspace*{2 mm}
\noindent
\makebox[16mm][l]{Ising} --- Ising antiferromagnet \\
\makebox[16mm][l]{HEP} --- Heisenberg triangular antiferromagnet with
Easy-Plane type anisotropy \\
\makebox[16mm][l]{HEA} --- Heisenberg triangular antiferromagnet with
Easy-Axis type anisotropy \\
\makebox[16mm][l]{SGS} --- Singlet-Ground-State magnet \\
\makebox[16mm][l]{F} --- triangular antiferromagnet stacked Ferromagnetically
\\
\makebox[16mm][l]{non Tr} --- magnetic structure is non Triangular \\
\makebox[16mm][l]{non Hex} --- crystal structure is non Hexagonal \\
\makebox[16mm][l]{$^\star$} --- CsNiI$_3$ is reported to be not a localized
spin system, but an itinerant electron system~\cite{rnc_rvb_cni_Tanaka_92}

\vspace*{5 mm}
\caption[]{Magnetic structure of ABX$_3$ triangular compounds}
\vspace*{5 mm}
\label{listofmagnstr}
\end{table}

\newpage
\small
\begin{table}[p] 
\vspace*{1mm}
\noindent
\begin{tabular}{clll}
\small 
            &  \multicolumn{1}{c}{$J$, GHz}      
            &  \multicolumn{1}{c}{$J'$, GHz}             
            &  \multicolumn{1}{c}{$D$, GHz}             \\*
            &
            & \multicolumn{1}{c}{or $J'$ and $J'_1$}
            &                             \\*\hline
CsNiCl$_3$  &  345(8)--NS\cite{cnc_Morra_88}     
            &  6.0(6)--NS\cite{cnc_Morra_88}     
            & -13.0(1.5)--NS\cite{cnc_Morra_88}         \\*
S=1         &  275--M\cite{cnc_Katori_95}
            &  8(1)--ESR\cite{cnc_Zaliznyak_89}
            &  -1.2--ESR\cite{cnc_Zaliznyak_89}         \\*
            &
            &  5.4--NS\cite{abx_Kakurai_92}
	    &                                           \\* 
RbNiCl$_3$  &  485--NS\cite{rnc_Tun_91}
            &  14--NS\cite{rnc_Tun_91}     
            &  -1.5--ESR\cite{rnc_Cohen_77}             \\*
S=1         &  496--NS\cite{rnc_Nakajima_92}
            &  38(4)--ESR\cite{rnc_Petrenko_90}
            &  -1.0(1)--ESR\cite{rnc_Petrenko_90}       \\*
RbNiBr$_3$  &  520--ESR\cite{rnb_Iio_87}
            &  21--M\cite{abx_Witteveen_74}
            &   ---                                     \\*  
CsNiBr$_3$ &  354--M\cite{cnb_Brener_77}
           &  6.5--NMR\cite{cnb_Maegawa_91}
           &  -13.5--NMR\cite{cnb_Maegawa_91}           \\*
S=1        &
           & 
           &  -31.2--M\cite{cnb_Brener_77}              \\*
CsMnI$_3$  &  198(2)--NS\cite{cmi_Harrison_91}
           &  1.0(1)--NS\cite{cmi_Harrison_91}
           &  -0.50(2)--NS\cite{cmi_Harrison_91}        \\*
S=5/2      & 
           &  0.9--NS\cite{cmi_Inami_94}
           &  -1.7--NS\cite{cmi_Tun_94}                 \\*
           &
           &  0.88--NS\cite{cmi_cnc_Enderle_93}
           &  -1.07--ESR\cite{cmi_Abarzhi_93}           \\*
CsMnBr$_3$ &  211(2)--NS\cite{cmb_Collins_84}   
           &  0.46(5)--NS\cite{cmb_Gaulin_87}    
           &  3.4(5)--NS\cite{cmb_Gaulin_87}            \\*
S=5/2      &  215(3)--NS\cite{cmb_Falk_87}
	   &  0.41(2)--NS\cite{cmb_Falk_87}
	   &  2.9(3)--NS\cite{cmb_Falk_87}              \\*
           &
           &  0.46--NS\cite{abx_Inami_95}
           &  2.9--NS\cite{abx_Inami_95}                \\* 
RbMnBr$_3^\star$ &  199--NS\cite{rmb_Heller_94}   
           &  0.54--NS\cite{rmb_Heller_94}    
           &  2.2--NS\cite{rmb_Heller_94}               \\* 
S=5/2      &  186--M\cite{rmb_Abanov_94}
           &  0.22--M\cite{rmb_Abanov_94}
           &  1.3--M\cite{rmb_Abanov_94}                \\*
KNiCl$_3^\star$  &  310(6)--NS\cite{knc_Petrenko_95}   
           &      
           &  130(10)--NS\cite{knc_Petrenko_95}         \\*
S=1        &  312--M\cite{knc_Tanaka_89}
           &  \multicolumn{1}{c}{0.23 and 0.27--ESR\cite{knc_Tanaka_89}}
           &  18--M\cite{knc_Tanaka_89}                 \\*
CsVBr$_3$  &  1700--NS\cite{avx_Niel_77}   
           &  0.43--NS\cite{cvc_cvb_cbi_Feile_84}    
           &  0.48--NS\cite{cvc_cvb_cbi_Feile_84}       \\*
S=3/2      &  1700-1900--M\cite{avx_Niel_77}
           &  
           &                                            \\*
RbVBr$_3$  &  2700--NS\cite{cvc_Kakurai_9?}       
           &  \multicolumn{1}{c}{$J'$ and 1.7$J'$--NS\cite{rvb_Tanaka_94}} 
           &     ---                                     \\*
S=3/2      &  
           &
           &                                             \\*     
CsVCl$_3$  &  2400--NS\cite{avx_Niel_77}   
           &  0.15--NS\cite{cvc_cvb_cbi_Feile_84}    
           &  0.29--NS\cite{cvc_cvb_cbi_Feile_84}        \\*
S=3/2      &  3500--NS\cite{cvc_Kadowaki_83}
           &  1.0--NS\cite{cvc_Kadowaki_83}
           &                                             \\*
           &  2700--NS\cite{cvc_Itoh_95}
           &  0.8--NS\cite{abx_Kakurai_92}
           &                                             \\*
           &  2400--M\cite{avx_Niel_77}
           &
           &                                             \\*
CsVI$_3$   &  1100--NS\cite{avx_Niel_77}   
           &  1.9--NS\cite{cvc_cvb_cbi_Feile_84}    
           &  3.4--NS\cite{cvc_cvb_cbi_Feile_84}         \\*
S=3/2      &  1100-1400--M\cite{avx_Niel_77}
           &
           &                                             \\*
VCl$_2$    &  2.8(1)--NS\cite{vcl2_Kadowaki_87}
           &  458(13)--NS\cite{vcl2_Kadowaki_87}
           &  -1.5--NS\cite{vcl2_Kadowaki_87}            \\*
 S=3/2     &       
           &  480--M\cite{cvx_vx2_Niel_77}
           &  -2--ESR\cite{vx2_Yamada_84}                \\*
VBr$_2$    & 4--NS\cite{vbr2_Kadowaki_85}
           & 333--NS\cite{vbr2_Kadowaki_85}     
           & -2--ESR\cite{vx2_Yamada_84}                 \\*
S=3/2      &    
           & 333--M\cite{cvx_vx2_Niel_77}
           &                                             \\*
VI$_2$     &   ---     
           & 125--M\cite{cvx_vx2_Niel_77}
           &     ---                                     \\*
  S=3/2    &
           &
           &                                             \\*
LiCrO$_2$ & --- 
                   & 810--M\cite{abo_Delmas_78} 
                   &  ---                          \\*
S=3/2       & 
                  & 780--ESR\cite{abo_Angelov_84}
                  &                                  \\*
CuCrO$_2$ & ---
                   & 236--M\cite{abo_Doumerc_86} 
                   &    ---                        \\*
S=3/2       &
                  &
                  &                                  \\*
AgCrO$_2$ & ---
                   & 186--M\cite{abo_Doumerc_86}
                   & ---                           \\*
S=3/2       &
                  &
                  &                            \\*
\end{tabular} 
\vspace*{3 mm}
\noindent
\makebox[16mm][l]{NS} --- Neutron Scattering measurements\\*
\makebox[16mm][l]{ESR} --- Electron Spin Resonance measurements\\*
\makebox[16mm][l]{NMR} --- Nuclear Magnetic Resonance measurements\\*
\makebox[16mm][l]{M} --- Magnetization and susceptibility measurements \\*

\noindent
In the literature a variety of units are used for quantities listed in the 
table. The conversion factors are \newline
1 GHz = 4.136 $\mu$eV =0.0480 K = 0.03336 cm$^{-1}$.

\vspace*{1mm}

\noindent
$^\star$ Data for RbMnBr$_3$ and KNiCl$_3$ refer to orthorhombic and hexagonal
phases respectively 

\vspace*{3 mm}
\caption[]{Exchange and anisotropy constants for triangular Heisenberg 
antiferromagnets.}
\vspace*{3 mm}
\label{Heis_parameters}
\end{table}

\newpage

\begin{table}[p]
\vspace*{1mm}
\noindent
\begin{tabular}{cccccc}
Exponent        &  VCl$_2$
        &  VBr$_2$
        &  SO(3)\cite{Plumer}
        &  Tricritical
        &  Heisenberg\cite{Plumer}  \\*  \hline
$\beta$ &  0.20(2)\cite{vcl2_Kadowaki_87}
        &  0.20\cite{vbr_Kadowaki_85}
        &  0.30(2)
        &  0.25
        &  0.368(4)    \\*
$\nu$   &  0.62(5)\cite{vcl2_Kadowaki_87}
        & ---
        &  0.59(2)
        &  0.5
        &  0.710(7)        \\*
$\gamma$        &  1.05(3)\cite{vcl2_Kadowaki_87}
        &  ---
         &  1.17(2)
        &  1.0
        &  1.390(10)       \\*
$\alpha$        & ---
        &  0.30(5)\cite{vbr2_Wosnitza_94}
        &  0.24(8)
        &  0.5
        &  -0.126(11)        \\*
        &
        &  0.59(5) and 0.28(2)\cite{vbr2_Takeda_86}
        &
        &
  
        &        \\*
\end{tabular}

\vspace*{3 mm}
\caption[]{Experimental values of critical exponents for frustrated Heisenberg 
systems, compared with three models.}
\vspace*{3 mm}
\label{critexp}
\end{table}

\begin{table}
\vspace*{3cm}
\noindent
\begin{tabular}{clllllr}
          & \multicolumn{1}{c}{T$_{N1}$, T$_{N2}$, K}
          & \multicolumn{1}{c}{T$_M$, K}
          & \multicolumn{1}{c}{H$_C$, T}
          & \multicolumn{1}{c}{H$_M$, T}
          & \multicolumn{1}{c}{{\Large $\theta$}$_{T=0}$, $^\circ$}
&\multicolumn{1}{c}{{\Large $\mu$}$_{T=0}$, {\Large $\mu$}$_B$} \\*[1mm] \hline
CsNiCl$_3$  &  4.84, 4.40--NMR\cite{cnc_Clark_72}
            &  4.48--C\cite{cnc_Beckmann_93_2}
            &  1.99--M\cite{cnc_rnc_Johnson_79}
            &  2.25--C\cite{cnc_Beckmann_93_2}
            &  59--NS\cite{cnc_Kadowaki_87}
            &  1.1(1)--NS\cite{cnc_Cox_71,cnc_Yelon_73}          \\*
S=1         &  4.88(5), 4.40(5)--M\cite{cnc_rnc_Johnson_79}
            &  4.495--US\cite{cnc_Poirier_90}
            &  1.9--ESR\cite{cnc_Zaliznyak_88,cnc_Zaliznyak_89}
            &  2.105--US\cite{cnc_Poirier_90}
            &  50--ESR\cite{abx_Tanaka_88}
            &  1.4(2)--NS\cite{rnc_cnc_Minkiewicz_70}          \\*
            &  4.83(8), 4.46(8)--NS\cite{cnc_Morra_88}
            &  4.48--ESR\cite{cnc_Poirier_90}
            &
            &  2.13--ESR\cite{cnc_Poirier_90}
            &  59-ESR\cite{cnb_Kambe_95}
            &                                  \\*
            &  4.80(1), 4.388(4)--US\cite{cnc_Trudeau_93}
            &
            &
            &
            &
            &  \\*[1mm]
RbNiCl$_3$  &  11.25, 11.11--NS\cite{rnc_Oohara_91_1}
            &  11.8--M\cite{cnc_rnc_Johnson_79}
            &  2.05--M\cite{cnc_rnc_Johnson_79}
            &  2.65--M\cite{cnc_rnc_Johnson_79}
            &  57.5--NS\cite{rnc_Yelon_72}
            &  1.3(1)--NS\cite{rnc_Yelon_72}                    \\*
S=1         &
            &
            &  2.01--ESR\cite{rnc_Petrenko_90}
            &  2.4--ESR\cite{rnc_Petrenko_90}
            &
            &  1.5(2)--NS\cite{rnc_cnc_Minkiewicz_70}          \\*[1mm]
CsNiBr$_3$ &  14.06, 11.51--NMR\cite{cnb_Maegawa_91}
           &  11.0--M\cite{cmi_cnb_Katori_93}
           &  8.8--M\cite{cmi_cnb_Katori_93}
           &  9.88--M\cite{cmi_cnb_Katori_93}
           &  39--NMR\cite{cnb_Maegawa_91}
           &  ---                                                 \\*
S=1        &  14.25, 11.75--C\cite{cnb_Brener_77}
           &
           &
           &
           &  58--ESR\cite{cnb_Kambe_95}
           &                                                      \\*
           &  13.46, 11.07--B\cite{cnx_Sano_89}
           &
           &
           &
           &
           &                                                     \\*[1mm]
RbNiBr$_3$ &  23.5, 21.47--ESR\cite{rnb_Iio_87}
           &  ---
           &  ---
           &  ---
           &  ---
           &  ---                                      \\*
S=1        &
           &
           &
           &
           &
           &                                            \\*[1mm]
CsMnI$_3$  &  11.2, 8.17--NS\cite{cmi_Kadowaki_91}
           &  8.85--M\cite{cmi_cnb_Katori_93}
           &  5.3--M\cite{cmi_cnb_Katori_93}
           &  5.95--M\cite{cmi_cnb_Katori_93}
           &  50(2)--NS\cite{cmi_Zandbergen_80}
           &  3.7--NS\cite{cmi_Zandbergen_80}                  \\*
S=5/2      &  11.41, 8.21--NS\cite{cmi_Harrison_91}
           &  9.02(5)--B\cite{cmi_cnc_Enderle_94}
           &  5.4--M\cite{cmi_Zandbergen_80}
           &  5.86(1)--B\cite{cmi_cnc_Enderle_94}
           &  51(1)--NS\cite{cmi_Harrison_91}
           &                                                \\*
           &  11.20(1) 8.166(5)--NS\cite{cmi_Ajiro_90}
           &
           &  5.2--ESR\cite{cmi_Abarzhi_93}
           &
           & 55-ESR\cite{cnb_Kambe_95}
           &                                               \\*
\end{tabular}
\vspace*{3 mm}
\noindent
\makebox[10mm][l]{NS} --- Neutron Scattering measurements\\*
\makebox[10mm][l]{ESR} --- Electron Spin Resonance measurements\\*
\makebox[10mm][l]{NMR} --- Nuclear Magnetic Resonance measurements\\*
\makebox[10mm][l]{M} --- Magnetization and susceptibility measurements \\*
\makebox[10mm][l]{C} --- Specific Heat measurements\\*
\makebox[10mm][l]{US} --- Ultrasonic velocity and attenuation measurements \\*
\makebox[10mm][l]{B} --- Birefringence measurements \\*

\vspace*{3 mm}
\caption[]{Characteristics of the Heisenberg triangular
antiferromagnets with Easy-Axis anisotropy.}
\vspace*{3 mm}
\label{HEA_tab}
\end{table}

\newpage
\begin{table}[p]
\begin{tabular}{ccccccc}
Exponent        & Material
        & Experimental
        & Chiral
        & Chiral
        &
        &                   \\*
        &
        & value
        & XY
        & Heisenberg
        & XY
        & Heisenberg      \\*[1mm] \hline
$\beta _1$      & CsNiCl$_3$
                & 0.32(3) \cite{cnc_Clark_72}
        & 0.25(1)
        & 0.30(2)
        & 0.35
        & 0.36             \\*
        & CsNiCl$_3$
        & 0.30(2)  \cite{cnc_Kadowaki_87}
        &
        &
        &
        &     \\*
        & RbNiCl$_3$
        & 0.27(1) \cite{rnc_Oohara_91_1}
        &
        &
        &
        &     \\*
        & CsMnI$_3$
        & 0.32(1) \cite{cmi_Ajiro_90}
        &
        &
        &
        &     \\*
$\nu _1$        & CsMnI$_3$
        & 0.59(3) \cite{cmi_Kadowaki_91}
        & 0.54(2)
        & 0.59(2)
        & 0.669
        & 0.705   \\*
$\gamma _1$     & CsMnI$_3$
        & 1.12(7) \cite{cmi_Kadowaki_91}
        & 1.13(5)
        & 1.17(7)
        & 1.316
        & 1.387     \\*
$\alpha _1$     & CsNiCl$_3$
        & -0.05(8) \cite{rev_Weber_95}
        & 0.34(6)
        & 0.24(8)
        & -0.008
        & -0.116   \\*
(A$^+$/A$^-$)$_1$       & CsNiCl$_3$
        & 1.21(5) \cite{rev_Weber_95}
        & 0.36(2)
        & 0.54(2)
        & 0.99
        & 1.36    \\*
        &       &       &       &       &       &\\*
$\beta _2$      & CsNiCl$_3$
                & 0.32(3) \cite{cnc_Clark_72}
        & 0.25(1)
        & 0.30(2)
        & 0.35
        & 0.36             \\*
        & CsNiCl$_3$
        & 0.30(2)  \cite{cnc_Kadowaki_87}
        &
        &
        &
        &     \\*
        & RbNiCl$_3$
        & 0.28(1) \cite{rnc_Oohara_91_1}
        &
        &
        &
        &     \\*
        & CsMnI$_3$
        & 0.35(1) \cite{cmi_Ajiro_90}
        &
        &
        &
        &     \\*
$\nu _2$        & CsMnI$_3$
        & 0.56(2) \cite{cmi_Kadowaki_91}
        & 0.54(2)
        & 0.59(2)
        & 0.669
        & 0.705   \\*
$\gamma _2$     & CsMnI$_3$
        & 1.04(3) \cite{cmi_Kadowaki_91}
        & 1.13(5)
        & 1.17(7)
        & 1.316
        & 1.387     \\*
$\alpha _2$     & CsNiCl$_3$
        & -0.06(10) \cite{rev_Weber_95}
        & 0.34(6)
        & 0.24(8)
        & -0.008
        & -0.116   \\*
        & CsMnI$_3$
        & -0.05(15) \cite{rev_Weber_95}
        &
        &
        &
        &        \\*
(A$^+$/A$^-$)$_2$       & CsNiCl$_3$
        & 1.2(3) \cite{rev_Weber_95}
        & 0.36(2)
        & 0.54(2)
        & 0.99
        & 1.36    \\*
        & CsMnI$_3$
        & 1.2     \cite{rev_Weber_95}
        &
        &
        &
        &     \\*
        &       &       &       &       &       &\\*
$\beta_M$       &CsNiCl$_3$
        &0.28(3)\cite{cnc_Enderle_97}
        &0.25(1)
        &0.30(1)
        &0.35
        &0.36   \\*
 $\alpha _M$    & CsNiCl$_3$
        & 0.25(8) \cite{rev_Weber_95}
        & 0.34(6)
        & 0.24(8)
        & -0.008
        & -0.116   \\*
        & CsNiCl$_3$
        & 0.23(4) \cite{cmi_cnc_Enderle_94}
        &
        &
        &
        &        \\*
        & CsMnI$_3$
        & 0.28(6) \cite{rev_Weber_95}
        &
        &
        &
        &   \\*
        & CsMnI$_3$
        & 0.44(3) \cite{cmi_cnc_Enderle_94}
        &
        &
        &
        &        \\*
(A$^+$/A$^-$)$_M$       & CsNiCl$_3$
        & 0.52(10) \cite{rev_Weber_95}
        & 0.36(2)
        & 0.54(2)
        & 0.99
        & 1.36    \\*
        & CsMnI$_3$
        & 0.42(10) \cite{rev_Weber_95}
        &
        &
        &
        &     \\*
        &       &       &       &       &       &\\*
$\beta_F$               &CsNiCl$_3$
        &0.243\cite{cnc_Enderle_97}
        &0.25(1)
        &0.30(2)
        &0.35
        &0.36   \\*
$\alpha _F$     & CsNiCl$_3$
        & 0.37(8) \cite{rev_Weber_95}
        & 0.34(6)
        & 0.24(8)
        & -0.008
        & -0.116   \\*
        & CsNiCl$_3$
        & 0.342(5) \cite{cmi_cnc_Enderle_94}
        &
        &
        &
        &        \\*
        & CsMnI$_3$
        & 0.34(6) \cite{rev_Weber_95}
        &
        &
        &
        &   \\*
(A$^+$/A$^-$)$_F$       & CsNiCl$_3$
        & 0.30(11) \cite{rev_Weber_95}
        & 0.36(2)
        & 0.54(2)
        & 0.99
        & 1.36    \\*
        & CsMnI$_3$
        & 0.31(8) \cite{rev_Weber_95}
        &
        &
        &
        &     \\*
\end{tabular}

\vspace*{3 mm}
\caption[]{Observed critical exponents for easy-axis materials
and predicted critical exponents~\cite{cmi_cnc_Enderle_94,rev_Weber_95}
for various universality classes. Subscripts 1, 2, $M$ and $F$ represent
exponents at $T_{N1}$, $T_{N2}$, the multicritical point and between the
spin-flop and the paramagnetic phase respectively.}
\vspace*{3 mm}
\label{hea_crit}
\end{table}

\newpage
\begin{table}[p]
\vspace*{1mm}
\begin{tabular}{cllllr}
           &  \multicolumn{1}{c}{Ordering}
                 &  \multicolumn{1}{c}{Space Group}
           &  \multicolumn{1}{c}{$T_N$, K or }
           &  \multicolumn{1}{c}{$H_C$, T}
           &  \multicolumn{1}{c}{{\Large $\mu$}$_{T=0}$, {\Large $\mu$}$_B$ }\\
           &  \multicolumn{1}{c}{type}
                 &  \multicolumn{1}{c}{at low T}
           &  \multicolumn{1}{c}{$T_{N1}$ and $T_{N2}$, K}
           &
           &                                   \\ \hline
CsMnBr$_3$ &  $D>3J'$
                    &  $P6_3/mmc$
                    &  8.32--NS\cite{cmb_Gaulin_89}
                    &  6.2--NS\cite{cmb_Gaulin_89}
                    &  3.3--NS\cite{cmb_Eibshutz_72}                         \\
 S=5/2         &
           &
           &
           &
           &                                                        \\
CsVBr$_3$  &  $D<3J'$
           &  $P6_3/mmc$
           &  20.4--NS\cite{avx_Hauser_85}
           &  ---
           &  1.87--NS\cite{avx_Hauser_85}                          \\
 S=3/2     &
           &
           &  20.3--M\cite{cvb_rvb_Tanaka_94}
           &
           &            \\
CsVCl$_3$  &  $D<3J'$
           &  $P6_3/mmc$
           &  13.8--NS\cite{avx_Hauser_85}
           &  ---
           &  1.97--NS\cite{avx_Hauser_85}   \\
 S=3/2     &
           &
           &
           &
           &            \\
RbVCl$_3$  &  $D<3J'$
           &  $P6_3/mmc$
           &  19(1)--NS\cite{avx_Hauser_85}
           &  ---
           &  2.31--NS\cite{avx_Hauser_85} \\
 S=3/2     &
           &
           &
           &
           &            \\
CsVI$_3$   &  $D<3J'$
           &  $P6_3/mmc$
           &  34.8--NS\cite{avx_Hauser_85}
           &  ---
           &  1.64--NS\cite{avx_Hauser_85}           \\
 S=3/2     &
           &
           &  32(1)--NS\cite{abi_Zandbergen_81}
           &
           &            \\
RbMnBr$_3$ &  Distorted
           &  there are two phases:
           &
           &
           &                            \\
   S=5/2   &
           &  hexagonal ($\leq P6_3/mmc$)
           &  10.0--NS\cite{rmb_Heller_94}
           &  ---
           &  ---                       \\
           &
           &  orthorhombic ($Pbcm$ or $Pca2_1$)
           &  8.5--NS\cite{rmb_Heller_94}
           &  3.9--M\cite{rmb_Bazhan_93}, 4.0--NS\cite{rmb_Kato_93}$^\star$
           &  3.6--NS\cite{rmb_Glinka_73}                        \\
KNiCl$_3$  &  Distorted
           &  there are two phases:
           &
           &
           &  \\
  S=1      &
           &  hexagonal ($\leq P6_3/mmc$)
           &  8.6--M\cite{knc_Petrenko_95}, NS\cite{knc_Petrenko_96a}
           &  2.3--M\cite{knc_Petrenko_95}, 1.8--ESR\cite{knc_Tanaka_89}
           &  ---                                                   \\
           &
           &  orthorhombic ($Pbcm$ or $Pca2_1$)
           &  12.5--NS\cite{knc_Petrenko_96a}
           &  ---
           &  ---                                          \\
RbVBr$_3$  &  Distorted
           &  $P6_3cm$ or $P\overline{3}c1$
           &  28.1 and 21.0--M\cite{cvb_rvb_Tanaka_94}
           &  ---
           &  1.53--NS\cite{avx_Hauser_85}                             \\
 S=3/2     &
           &
           &
           &
           &            \\
RbVI$_3$   &  Distorted
           &  $P6_3cm$ or $P\overline{3}c1$
           &  25--NS\cite{avx_Hauser_85}
           &  ---
           &  1.44-NS\cite{avx_Hauser_85} \\
 S=3/2     &
           &
           &
           &
           &            \\
RbTiI$_3$  & Distorted
           & $P6_3cm$ or $P\overline{3}c1$
           & $<4.2$-NS\cite{abi_Zandbergen_81}
           & ---
           & ---            \\
  S=1      &
           &
           &
           &
           &              \\
\end{tabular}

\vspace*{3 mm}
$^\star$ In orthorhombic phase of RbMnBr$_3$ beside transition from
triangular (or close to triangular) to collinear magnetic structure
at $H_C$, there is incommensurate-commensurate phase transition at
$H \approx 3$~T. For details see part~\ref{Distort}.  

\caption[]{Characteristics of the Heisenberg triangular antiferromagnets 
with Easy-Plane type anisotropy.}
\vspace*{3 mm}
\label{HEP_table}
\end{table}

\newpage
\begin{table}
\vspace*{1mm}
\begin{tabular}{cccccc}
Exponent        & Experimental
        & Chiral
        & Chiral
        & XY
        & Mean Field   \\
        & value
        & XY
        & Heisenberg
        &
        & Tricritical  \\
        &       &       &       &       &  \\        \hline
$\beta$ & 0.22(2) \cite{cmb_Mason_87}
        & 0.25(1)
        & 0.30(2)
        & 0.35
        & 0.25   \\
        & 0.25(1) \cite{cmb_Ajiro_88}
        &       &       &       &  \\
        & 0.21(2) \cite{cmb_Mason_89}
        &       &       &       &  \\
        & 0.24(2) \cite{cmb_Gaulin_89}
        &       &       &       &  \\
        & 0.28(2) \cite{rmb_Kato_95} $^*$
        &       &       &       &  \\
        &       &       &       &       &  \\
$\nu$   & 0.57(3) \cite{cmb_Kadowaki_88}
        & 0.54(2)
        & 0.59(2)
        & 0.669
        & 0.50     \\
        & 0.54(3) \cite{cmb_Mason_89}
        &       &       &       &  \\
        &       &       &       &       &  \\
$\gamma$        & 1.10(5) \cite{cmb_Kadowaki_88}
        & 1.13(5)
        & 1.17(7)
        & 1.316
        & 1.00     \\
        & 1.01(8) \cite{cmb_Mason_89}
        &       &       &       &  \\
        &       &       &       &       &  \\
$\alpha$        & 0.39(9) \cite{cmb_Wang_91}
        & 0.34(6)
        & 0.24(8)
        & -0.008
        & 0.50    \\
        & 0.40(5) \cite{cmb_Deutschmann_92}
        &       &       &       &  \\
        &       &       &       &       &  \\
$A^{+}/A^{-}$   & 0.19(10) \cite{cmb_Wang_91}
        & 0.36(2)
        & 0.54(2)
        & 0.99
        &  -    \\
        & 0.32(20) \cite{cmb_Deutschmann_92}
        &       &       &       &  \\
        &       &       &       &       &  \\
$\overline{\phi}$       & 0.98(7) \cite{cmb_Gaulin_89}
        & $1<\overline{\phi}<1.13$
        & $1<\overline{\phi}<1.17$
        & $1<\overline{\phi}<1.32$
        &  \\
        & 1.05(5) \cite{cmb_Goto_90}
        &       &       &       &  \\
        & 0.78(10) \cite{rev_Weber_95}
        &       &       &       &  \\
        & 0.79(6) \cite{cvb_rvb_Tanaka_94} $^\dagger$
        &       &       &       &  \\
        &       &       &       &       &  \\
z       & 1.46(6) \cite{cmb_Mason92}
        &  -    &  -    & 1.50  &  -   \\
\end{tabular}

\vspace*{3 mm}
\caption[]{Experimental values of critical exponents in easy plane
materials compared with predictions from various 
models~\cite{rev_Weber_95,Gaulin_94}. Experimental values refer
to CsMnBr$_3$ except that $^*$ and $^\dagger$ refer to RbMnBr$_3$
and CsVBr$_3$ respectively.}
\vspace*{3 mm}
\label{critind}
\end{table}

\begin {table}
\begin{tabular}{cllccllr}
           & \multicolumn{1}{c}{$T_{N1}$}
           & \multicolumn{1}{c}{$T_{N3}$,$T_{N2}$}
           & \multicolumn{1}{c}{$H_{c1}$}
           & \multicolumn{1}{c}{$H_{c2}$}
           & \multicolumn{1}{c}{$J$, GHz}
           & \multicolumn{1}{c}{$J'$, GHz}
           & \multicolumn{1}{c}{{\large $\epsilon $}}    \\*  \hline
CsCoCl$_3$ &  20.82--NS\cite{ccoc_Mekata_78}
           &  5.5,13.5--NS\cite{ccoc_Mekata_78}
           &  33.0--M\cite{ccc_Amaya_90}
           &  44.6--M\cite{ccc_Amaya_90}
           &  1557(15)--NS\cite{ccc_Tellenbach_77}
           &  171--M\cite{ccoc_ccob_rcol_Hori_90}
           &  0.094(7)--NS\cite{ccc_Tellenbach_77}       \\*
           &  21.3--NS\cite{ccc_Yoshizawa_79}
           &  9.2($T_{N3}$)--NS\cite{ccc_Yoshizawa_79}
           &
           &
           &  1541(12)--NS\cite{ccc_Yoshizawa_81}
           &  156--M\cite{ccl_Hori_92}
           &  0.14(2)--NS\cite{ccc_Yoshizawa_81}           \\*
           &  21--M\"{o}\cite{ccc_Ward_87}
           &  8.5--M\"{o}\cite{ccc_Ward_87}
           &
           &
           &  1495(10)--NS\cite{ccb_Nagler_83}
           &  30(2)--NS\cite{ccb_Nagler_83}
           &  0.120(3)--NS\cite{ccb_Nagler_83}         \\*
           &  21.01--NS\cite{ccoc_d_Mekata_90}
           &
           &
           &
           &  1676--M\cite{ccc_Amaya_90}
           &
           &  0.097--M\cite{ccc_Amaya_90}                 \\*
RbCoCl$_3$ &  28--M\"{o}\cite{ccb_rcc_Bocquet_88}
           &
           &  37.8--M\cite{ccoc_ccob_rcol_Hori_90}
           &  50.2--M\cite{ccoc_ccob_rcol_Hori_90}
           &  1928--M\cite{ccoc_ccob_rcol_Hori_90}
           &  186--M\cite{ccoc_ccob_rcol_Hori_90}
           &  0.091--M\cite{ccoc_ccob_rcol_Hori_90}          \\*
           &  28--R\cite{rcc_Lockwood_83}
           &  11--R\cite{rcc_Lockwood_83}
           &
           &
           &  1500--R\cite{rcc_Lockwood_83}
           &  45--R\cite{rcc_Lockwood_83}
           &  0.1--R\cite{rcc_Lockwood_83}   \\*
CsCoBr$_3$ &  28.34(5)--M\cite{ccob_Yelon_75}
           &  12,16--NS\cite{ccob_Yelon_75}
           &  40.6--M\cite{ccoc_ccob_rcol_Hori_90}
           &  56.6--M\cite{ccoc_ccob_rcol_Hori_90}
           &  1621(7)--NS\cite{ccb_Nagler_83}
           &  96--NS\cite{ccb_Nagler_83}
           &  0.137(5)--NS\cite{ccob_Nagler_83}           \\*
           &  28.3--M\"{o}\cite{ccb_rcc_Bocquet_88}
           &  12--M\"{o}\cite{ccb_rcc_Bocquet_88}
           &
           &
           &  1630--M\cite{ccoc_ccob_rcol_Hori_90}
           &  211--M\cite{ccoc_ccob_rcol_Hori_90}
           &  0.106--M\cite{ccoc_ccob_rcol_Hori_90}       \\*
           &  28.3(1)--NS\cite{Rogge_95}
           &  12,16--NS\cite{ccb_Farkas_91}
           &
           &
           &
           &
           &  \\*
RbCoBr$_3$ &  36--NS\cite{rcb_rnc_Minkiewicz_71}
           &
           &
           &
           &
           &
           &
\end{tabular}
\vspace*{3mm}
\noindent
\makebox[6mm][l]{M}  -- Magnetization measurements  \\*
\makebox[6mm][l]{M\"{o}} -- M\"{o}ssbauer effect measurements  \\*
\makebox[6mm][l]{NS} -- Neutron scattering measurements    \\*
\makebox[6mm][l]{R}  -- Raman scattering measurements \\*

\vspace*{3 mm}
\caption[]{Neel temperatures, $T_{N1}$ and $T_{N2}$, in Kelvin, critical 
magnetic fields, 
$H_{c1}$ and $H_{c2}$, in Tesla, exchange constants, $J$ and $J'$, in GHz,
and parameter $\epsilon$ for triangular Ising antiferromagnets. Values 
given for $\epsilon$ are for low temperature.}
\vspace*{3 mm}
\label{Ising_par}
\end{table}

\newpage
\begin{table}
\begin{tabular}{cccccc}
        &   Method
        &   $\alpha$
        &   $\beta$
        &   $\gamma$
        &   $\nu$         \\          \hline
        &
        &
        &
        &
        &                                                   \\
 3D XY model \cite{rev_Collins_89} &  Consensus
                                   &  -0.01(2)
                                   &  0.345(12)
                                   &  1.316(9)
                                   &  0.669(7)       \\
Matsubara \cite{Matsubara_87}   &  Monte Carlo
                                &
                                &  0.32(2)
                                &
                                &                    \\
Bunker {\it et al.} \cite{Bunker_93} &  Monte Carlo
                                                        &  -0.05(3)
                                                        &  0.311(4)
                                                        &  1.43(3)
                                                        &  0.685(3)  \\
Plumer, Mailhot \cite{PlumerM_95}       &  Monte Carlo
                                        &  0.012(30)
                                        &  0.341(4)
                                        &  1.31(3)
                                        &  0.662(9)      \\
                &
                &
                &
                &
                &                \\
Yelon {\it et al.} \cite{ccob_Yelon_75} &  Neutron
                                                        &
                                                        &  0.31(2)
                                                        &
                                                        &        \\
Mekata, Adachi \cite{ccoc_Mekata_78}    & Neutron
                                        &
                                        & 0.34(1)
                                        &
                                        &          \\
Mekata  {\it et al.} \cite{ccoc_d_Mekata_90}    &  Neutron
                                                        &
                                                        &  0.352
                                                        &
                                                        &     \\
Farkas {\it et al.} \cite{ccb_Farkas_91}        &  Neutron
                                                        &
                                                        &  0.22(2)
                                                        &
                                                        &       \\
Wang {\it et al.} \cite{ccb_Wang_94}    &  Spec. heat
                                                        &  -0.025(4)
                                                        &
                                                        &
                                                        &       \\
\end{tabular}

\vspace*{3 mm}
\caption[]{Comparision of different determinations of the critical indices for
the phase transition from an
ordered state to a paramagnetic state in Ising antiferromagnets}
\vspace*{3 mm}
\label{I_crit}
\end{table}

\newpage
\begin{table}
\noindent
\begin{tabular}{llll}
            & Space Group
            & \multicolumn{1}{c}{T$_N$, K}
            & Parameters of the magnetic interaction, GHZ     \\*
           & \multicolumn{1}{c}{at low T}
           &
           &                               \\* \hline
CsFeCl$_3$ & $P6_3/mmc$
           & LRO not found down to
           & $D=308, J=-148, J'=40$ (INS, heuristic formula)
             \cite{cfc_Steiner_81,cfc_Knop_83}                    \\*
          &
          & 0.8K (C) \cite{cfc_Takeda_80}, 80mK \cite{??_Dickson_81}
          & $D=523, J_\perp=-54.5, J'_\perp=2.88$ (INS, exciton model)
            \cite{cfc_rfc_Yoshizawa_80}                             \\*
         &
         &
         & $D=387, J_\parallel=-75, J_\perp=-150$ (M\"{o}, pair model)
           \cite{cfc_rfc_Montano_74}                                  \\*
        &
        &
        & $D=522, J=-62.9, J'=4.2$ (INS, RPA) \cite{cfc_Schmid_94}    \\*
       &
       &
       & $D=416, J_\parallel=-73, J_\perp=-110$ (NMR, spin-band model)
          \cite{cfc_Chiba_88}                                         \\*
      &
      &
      & $D=420, J=-78, J'=4.2$ (INS, DCEFA) \cite{cfc_Suzuki_95}      \\*
CsFeBr$_3$ & $P6_3/mmc$
           & LRO not found down to
           & $D=620, J=66, J'=6.7$ (INS, RPA) \cite{cfb_Dorner_88}  \\*
         &
         & 80 mK (NS) \cite{cfb_Schmid_92}
         & $D=620, J=66, J'=6.2$ (INS, RPA) \cite{cfb_Dorner_89} \\*
        &
        &
        & $D=640, J=64, J'=8$ (INS) \cite{cfb_Visser_91}     \\*
RbFeCl$_3$ & $P6_3/mmc$
           & 2.55 \cite{???_Davidson_71}
           & $D=360, J_\parallel=-150, J_\perp=-330$ (M\"{o}, pair model)
              \cite{cfc_rfc_Montano_74} \\*
          &
          & 2.45 (M\"{o}) \cite{cfc_rfc_Montano_74}
          & $D=408, J_\parallel=-110, J_\perp=-120, J'_\perp=16$  (INS,
            3 sublat. model) \cite{cfc_rfc_Yoshizawa_80}       \\*
        &
        &
        & $D=580, J_\perp=-65, J'_\perp=6.0$  (INS, exciton model)
          \cite{cfc_rfc_Yoshizawa_80}                            \\*
       &
       &
       & $D=498, J_\parallel=-30, J_\perp=-66$ (INS, DCEFA)
         \cite{rfc_Suzuki_81}                                     \\*
RbFeBr$_3$ & $P6_3cm$
           & 5.5 (NS) \cite{rfb_Eibschutz_73}
           & $D=250-270, J=52, J'=2$ (M\"{o}, CEFA) \cite{rfb_Lines_75} \\*
         &
         & 5.61 and 2.00 (C) \cite{rfb_Adachi_83}
         & $D=1580, J=26, J'=5.8$ (INS, SW-theory) \cite{rfb_Harrison_89}\\*
        &
        &
        & $D=1580, J=26, J'=5.8$ (INS, SW-theory) \cite{rfb_Harrison_89} \\*
       &
       &
       &                                                              \\*
\end{tabular}
\vspace*{3mm}
\noindent
\makebox[16mm][l]{LRO} --- Long Range Order \\
\makebox[16mm][l]{INS} --- Inelastic Neutron Scattering measurements\\
\makebox[16mm][l]{C} --- Specific Heat measurements \\
\makebox[16mm][l]{M\"{o}} --- M\"{o}ssbauer effect measurements \\
\makebox[16mm][l]{NMR} --- Nuclear Magnetic Resonance \\
\makebox[16mm][l]{RPA} --- Random Phase Approximation \\
\makebox[16mm][l]{DCEFA} --- Dynamical Correlated-Effective-Field
                             Approximation \\

\vspace*{3 mm}
\caption[]{Characteristics of the SGS triangular antiferromagnets}
\vspace*{3 mm}
\label{SGS_tab}
\end{table}

\begin{table}
\vspace{3mm}
\begin{tabular}{lllll}
 Experimental Technique & $-J$, GHZ
                & $J'$, GHz
                & $D$, GHz
                & $T_N$, K                         \\ \hline
 Neutron scattering & 239(2) \cite{cnf_Steiner_77}
                    &
                    & 185(4) \cite{cnf_Steiner_77}
                    & 2.67(5) \cite{cnf_Steiner_72}   \\*
                    &
                    &
                    &
                    &  2.664  \cite{cnf_Steiner_80}   \\*
 Specific heat      & 173(17) \cite{cnf_Lebesque_73}
                    &
                    &
                    & 2.613(3) \cite{cnf_Lebesque_73} \\*
 Magnetization      & 208(10) \cite{cnf_Dupas_77}
                    &
                    & 177(10) \cite{cnf_Dupas_77}
                    &                                 \\*
                    & 270 \cite{cnf_Scherer_77}
                    & 0.71 \cite{cnf_Scherer_77}
                    &
                    &                                 \\*
 AFMR               & 245 \cite{cnf_Yamazaki_79,cnf_Suzuki_83}
                    &
                    & 191(10) \cite{cnf_Yamazaki_79,cnf_Suzuki_83}
                    & 2.61 \cite{cnf_Yamazaki_79,cnf_Suzuki_83}  \\*
Ultrasonic velocity &
                    &
                    &
                    & 2.77(1)  \cite{cnf_Lussier_93}  \\
\end{tabular}
\vspace*{3 mm}
\caption[]{Various determinations of the magnetic parameters $J$, $J'$, $D$
and $T_N$ for CsNiF$_3$.}
\vspace*{3 mm}
\label{cnf}
\end{table}

\myfig{geom_frust}{Geometric frustration in the triangular
antiferromagnet}{\parbox[t]{13cm}{{\it a)} Geometric frustration arising from 
triangular arrangements of magnetic moments with each pair 
coupled antiferromagnetically. {\it b)} and {\it c)} Two degenerate 
solutions for the lowest energy of the system for a given spin vector 
at atom~1.}}

\myfig{cryst_str}{Crystal structure of ABX$_3$ and BX$_2$ compounds}
{\parbox[t]{13cm}{Crystal structure of ABX$_3$ (a) and BX$_2$ (b) compounds.
A is an alkali metal, B is a transition metal, and X is a halogen atom.}}

\myfig{stacked}{The stacked triangular antiferromagnet lattice}
{\parbox{13cm}{The stacked triangular antiferromagnet lattice.}} 

\myfig{HEA_PhD}{Magnetic phase diagram of a Heisenberg triangular
antiferromagnet with a small easy-axis anisotropy}{\parbox[t]{13cm}{Magnetic 
phase diagram of a Heisenberg triangular antiferromagnet with a small 
easy-axis anisotropy.}}

\myfig{HEP_smallD_PhD}{Magnetic phase diagram of a Heisenberg triangular
antiferromagnet with a small easy-plane anisotropy}{\parbox[t]{13cm}{Magnetic 
phase diagram of a Heisenberg triangular antiferromagnet with a small 
easy-plane anisotropy. There is both a tetracritical and a bicritical point.}}

\myfig{HEP_largeD_PhD}{Magnetic phase diagram of a Heisenberg triangular
antiferromagnet with large easy-plane anisotropy}{\parbox[t]{13cm}{Magnetic 
phase diagram of a Heisenberg triangular antiferromagnet with large 
easy-plane anisotropy. There is a tetracritical point at $T=T_N$ and $H=0$.}}

\myfig{CsMnBr3_int}{Temperature dependence of the intensity of magnetic
Bragg peaks in CsMnBr$_3$ at H$=4.2$~T}{\parbox[t]{13cm}{Temperature 
dependence of the intensity of magnetic Bragg peaks in CsMnBr$_3$ at $H=4.2$~T. 
Successive phase transitions from the paramagnetic phase to the spin-flopped 
phase and from the spin-flop to the triangular phase occur at 9.0~K and 
7.15~K respectively. The inset shows the field dependence of two nuclear 
Bragg peaks at T=7.0~K. Taken from Gaulin {\it et al.}~\cite{cmb_Gaulin_89}.}}

\myfig{CsMnBr3_M}{Magnetization of CsMnBr$_3$}{\parbox[t]{13cm}{The field 
(a) and temperature (b) dependence of the magnetization of CsMnBr$_3$.
Taken from Kotyuzhanskii and Nikivorov~\cite{cmb_Kotyuzhanskii_91} and from
Goto {\it et al.}~\cite{cmb_Goto_90}.}}

\myfig{knicl_dist}{Room temperature crystal structure of 
KNiCl$_3$}{\parbox[t]{13cm}{Room temperature crystal structure of KNiCl$_3$,
after Visser {\it et al.}~\cite{knc_Visser_80}. The two nickel-chlorine 
chains within a cell are displaced along the c axis relative to the chains 
at the cell corners.}}

\myfig{mag_dist}{Magnetic interactions on the distorted triangular 
lattice}{\parbox[t]{13cm}{Magnetic interactions on the distorted triangular 
lattice: a)~centered honeycomb model; b)~row model.}}

 \myfig{ising_spin_arr}{Basal plane ordering of the triangular Ising 
antiferromagnet}{\parbox[t]{13cm}{Some basal plane ordering of the triangular 
Ising antiferromagnet with unit cell $\sqrt{3a}$ by $\sqrt{3a}$, as marked 
by thick lines. Sites marked ``+'' have $S^z=\frac{1}{2}$, sites marked ``-''
have $S^z=-\frac{1}{2}$, sites marked ``O'' have $S^z$ randomly 
distributed with $<S^z>=0$, and sites marked $\frac{1}{2}$ have 
$S^z$ randomly distributed with $<S^z>=\frac{1}{4}$.}}

 \myfig{FYELON}{Temperature dependence of magnetic Bragg-peak intensity of
CsCoBr$_3$}{\parbox[t]{13cm}{Temperature dependence of some of the strong 
magnetic reflections in CsCoBr$_3$, after Yelon 
{\it et al.}~\cite{ccob_Yelon_75}. There are two critical phase 
transitions at $T_{N1}=28$~K and at $T_{N2}=12$~K.}}

 \myfig{FMEKATA}{Temperature dependence of magnetic Bragg-peak intensity of
CsCoCl$_3$}{\parbox[t]{13cm}{Temperature dependence of peak intensity of 
typical magnetic reflections of CsCoCl$_3$, after Mekata and 
Adachi~\cite{ccoc_Mekata_78}. Solid curves are calculated values.}}

 \myfig{ising_H_c}{Magnetization of CsCoCl$_3$}{\parbox[t]{13cm}{Magnetization 
process of CsCoCl$_3$ along the $c$-axis at several temperatures, after Amaya 
{\it et al.}~\cite{ccc_Amaya_90}. At low temperature there are phase       
transitions at $H_{c1}=33$~T and at $H_{c2}=45$~T.}}

\myfig{sgs_split}{Energy levels of Fe$^{2+}$ in the AFeX$_3$ 
family}{\parbox[t]{13cm}{Energy levels of Fe$^{2+}$ in the AFeX$_3$ family.}} 

\myfig{sgs_h_split}{Energy levels of SGS magnet in the magnetic
field}{\parbox[t]{13cm}{Energy level structure of the effective single-ion 
Hamiltonian [\ref{SGS_H}] for $H \parallel c$ (top) and $H \perp c$ (bottom).}}

\myfig{RbFeCl3_PhD}{Magnetic phase diagram of the RbFeCl$_3$}{\parbox[t]{13cm}
{Magnetic phase diagram of the RbFeCl$_3$ for $H \perp c$, after Wada {\it
et al.}~\cite{rfc_Wada_83}. Open and closed circles correspond to the anomalies 
observed in the specific heat and  susceptibility measurements respectively.}}

\myfig{CsNiF3}{Spin arrangement for CsNiF$3$.}{\parbox[t]{13cm}{Spin 
arrangement in the $ab$ plane for CsNiF$_3$. The unit cell is orthorhombic 
as shown by the solid lines.}}

\myfig{deluted}{Magnetic phase diagrams of CsNi$_{0.98}M_{0.02}$Cl$_3$}
{\parbox[t]{13cm}{Magnetic phase diagrams of CsNi$_{0.98}M_{0.02}$Cl$_3$, 
after Trudeau {\it et al.}~\cite{cnc_d_Trudeau_95}.}}

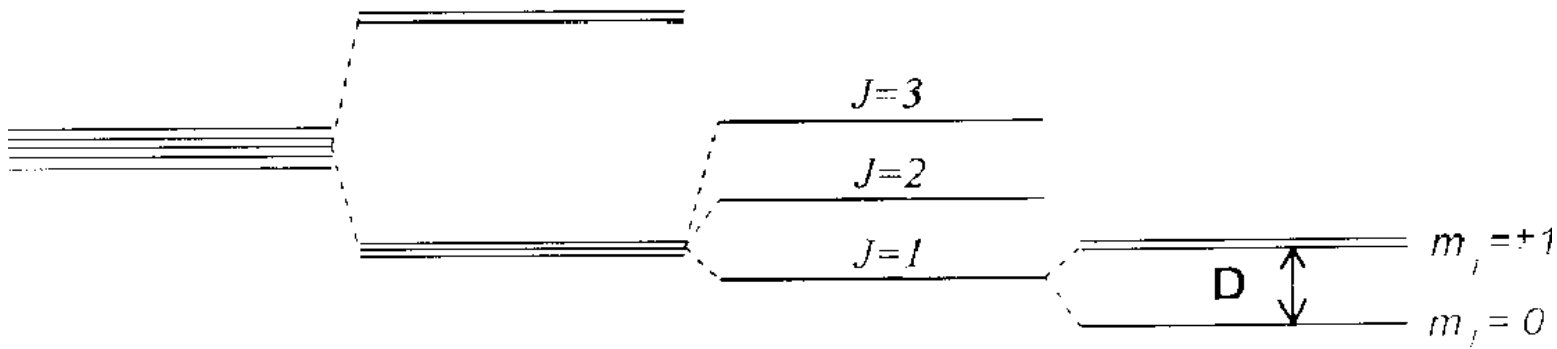
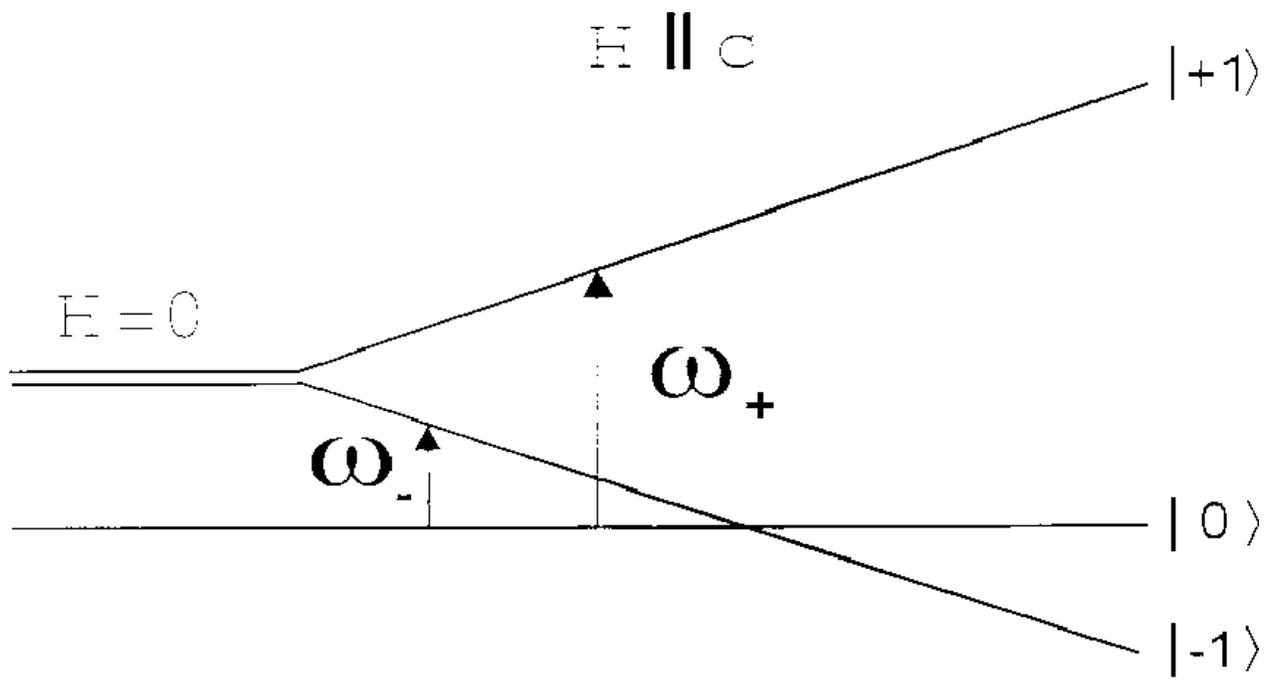
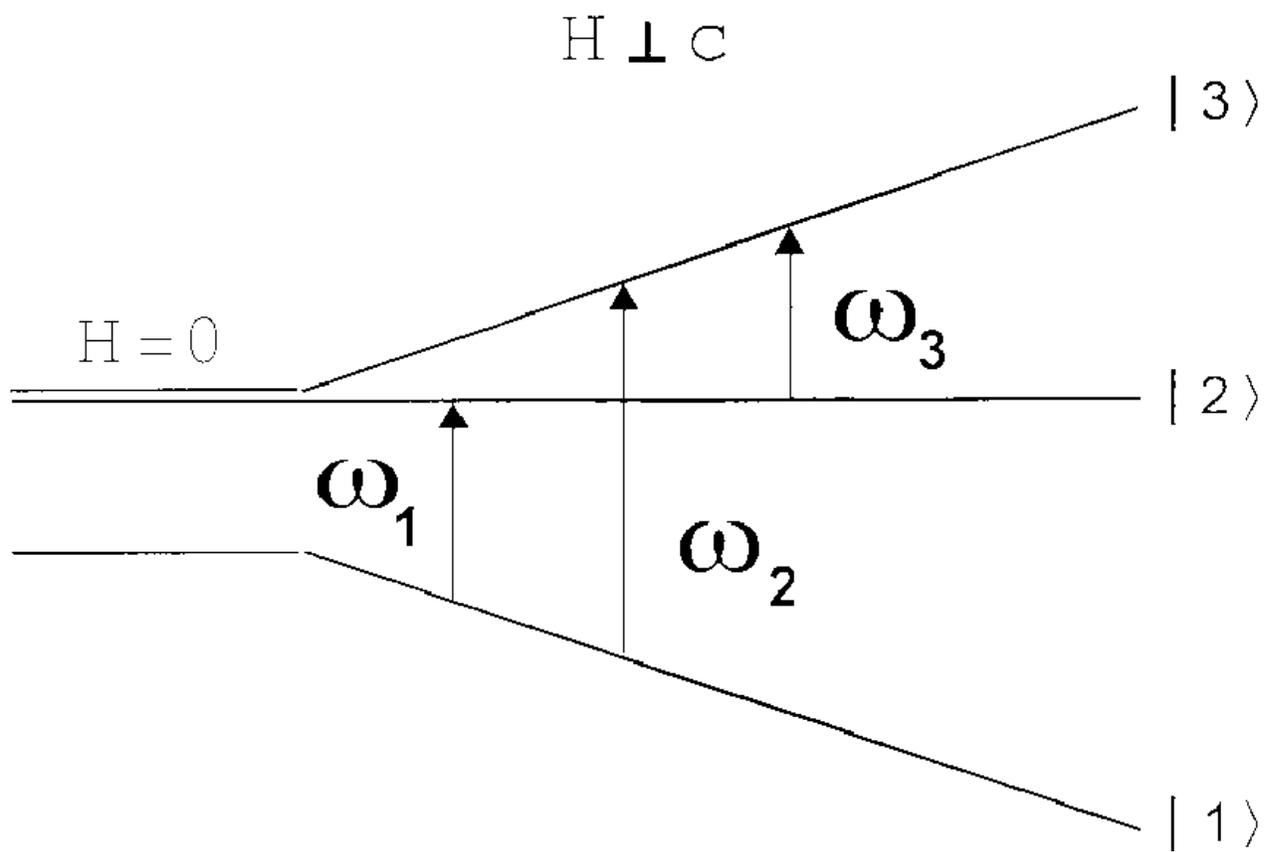
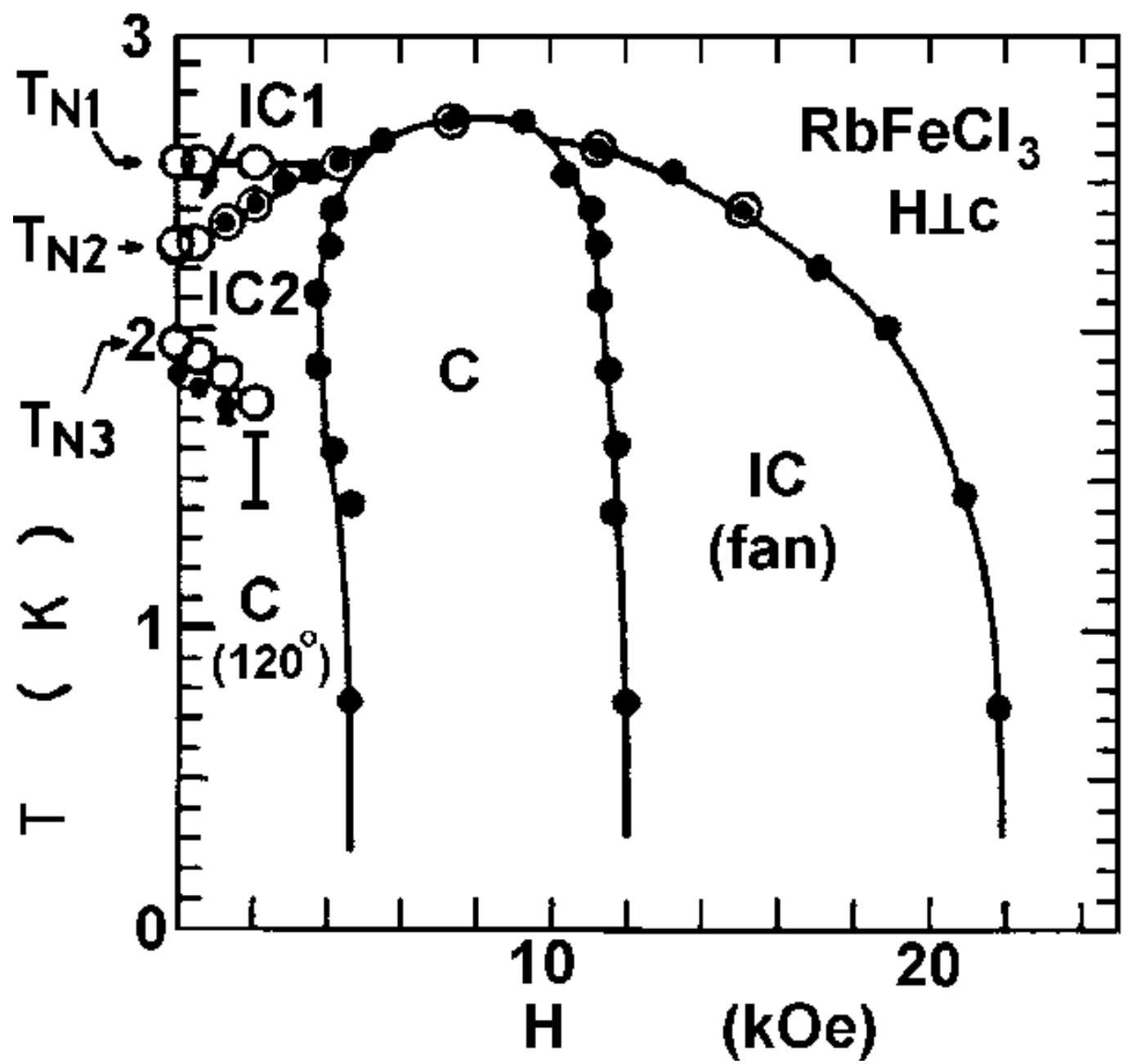
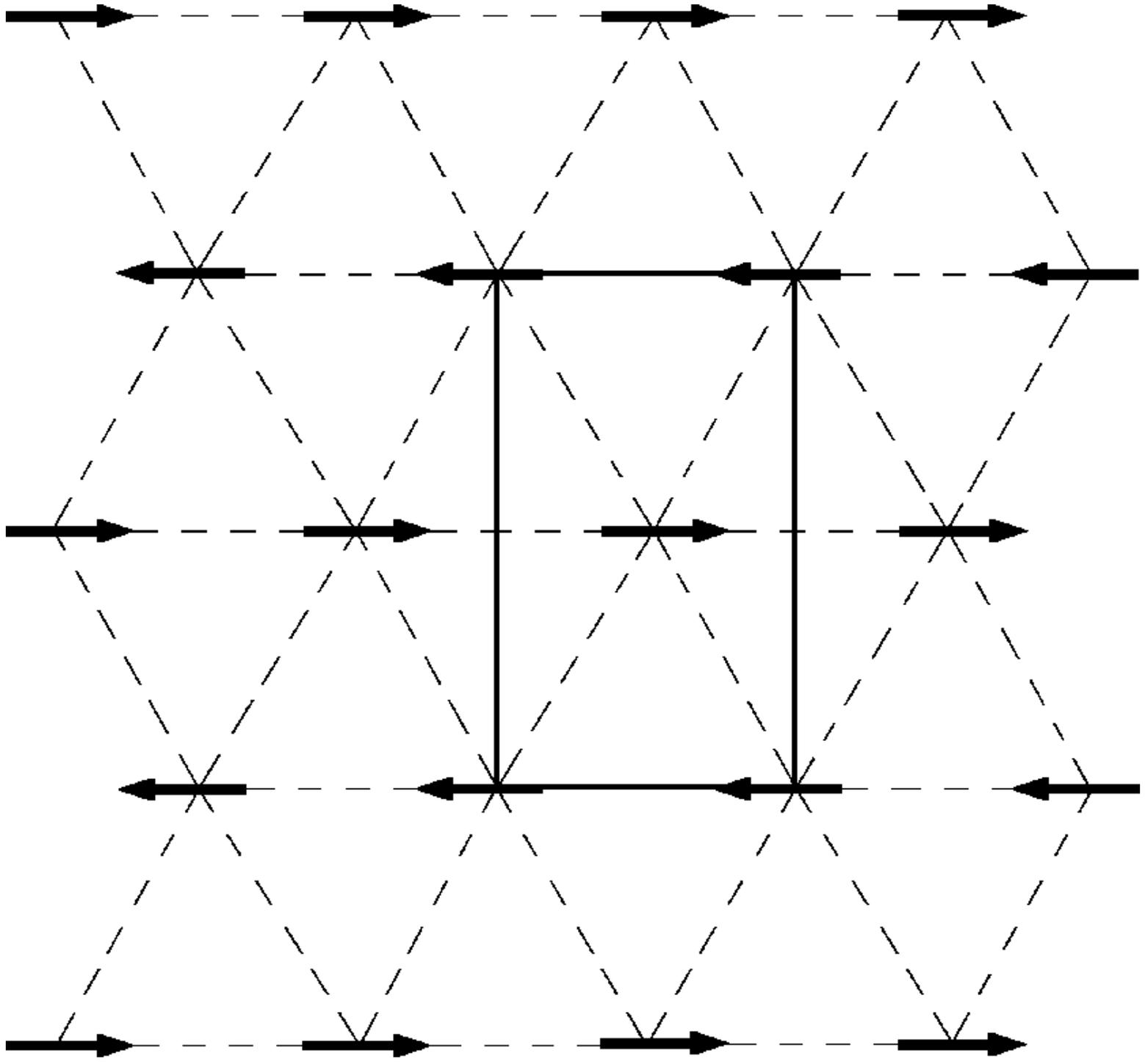
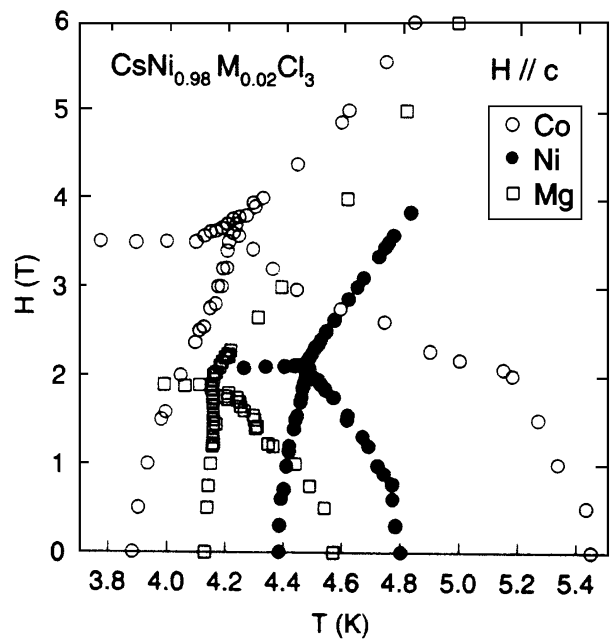
\end{document}